\documentclass[smallextended]{svjour3}
\usepackage[english]{babel}
\usepackage{color}
\usepackage{graphicx}   
\usepackage{amsmath}     
\usepackage{dingbat}
\usepackage{amssymb}
\usepackage[export]{adjustbox}
\usepackage{mdframed}
\usepackage{tabularx}
\usepackage{xcolor,colortbl}
\usepackage{pifont}
\usepackage{paralist}
\usepackage{makecell}
\usepackage{multirow}
\usepackage{amsfonts}
\usepackage{booktabs}
\usepackage{siunitx}
\usepackage{etoolbox}
\usepackage[utf8]{inputenc}
\usepackage[colorlinks,citecolor=blue,linkcolor=blue]{hyperref}
\usepackage{algorithm}
\usepackage[noend]{algpseudocode}
\usepackage{booktabs}
\usepackage{paralist}
\usepackage{listings,setspace,textcomp,chngcntr}
\usepackage{textcomp}
\usepackage{enumitem}
\usepackage{longtable}
\usepackage{numprint}
\usepackage{xspace}
\usepackage{mathtools}
\usepackage{footnote}
\usepackage{tcolorbox}
\usepackage{verbatimbox}
\usepackage{multirow}
\usepackage{adjustbox}
\usepackage{newfloat,caption}
\usepackage{float}
\usepackage{subcaption}
\usepackage{csquotes}
\usepackage[flushleft]{threeparttable}
\definecolor{bluekeywords}{rgb}{0.13,0.13,1}
\definecolor{greencomments}{rgb}{0,0.55,0.2}
\definecolor{redstrings}{rgb}{0.9,0,0}
\definecolor{lbcolor}{rgb}{0.9,0.9,0.9} 
\usepackage{placeins}

\usepackage[english]{babel}
\usepackage{natbib}

\setcitestyle{authoryear,open={(},close={)}}

\makeatletter
\newcommand\notsotiny{\@setfontsize\notsotiny\@vipt\@viipt}
\makeatother

\definecolor{darkgreen}{rgb}{0.01, 0.75, 0.24}
\newcommand{\TODO}[1]{\textcolor{red}{#1}}\newcommand\todo\TODO

\newcommand{\EMSErevision}[1]{\textcolor{black}{#1}}
\newcommand{\EMSErevisionMinor}[1]{\textcolor{black}{#1}}
\definecolor{tableShade}{gray}{0.9}

\newcommand\aseevogeneratedvalidtests{\numprint{141170}\xspace}

\newcommand\aserandoopgeneratedvalidtests{\numprint{14932884}\xspace}
\newcommand\emseevogeneratedvalidtests{\numprint{62513}\xspace}
\npdecimalsign{\ensuremath{\cdot}}\npthousandsep{,}
\newcommand\rgttotalnumber{\numprint{4477707}\xspace}

\DeclareFloatingEnvironment[fileext=frm,placement={!ht},name=Listing]{listing}

\usepackage{amssymb}
\usepackage{pifont}
\newcommand\RQone{To what extent does RGT patch assessment technique identify misclassified patches in previously reported research in program repair?}
\newcommand\RQtwo{To what extent does RGT patch assessment yield false positives?}
\newcommand\RQthree {To what extent is RGT patch assessment good at discarding overfitting patches compared against the state-of-the-art?}
\newcommand\RQfour{What is the time cost of RGT patch assessment?}
\newcommand\RQfive{What is the trade-off between test generation cost and patch classification effectiveness of RGT?}

\begin{document}

\title{Automated Patch Assessment for Program Repair at Scale}

\author{He Ye        \and
        Matias Martinez \and
        Martin Monperrus 
}

\institute{He Ye,  Martin Monperrus \at
              KTH Royal Institute of Technology, Sweden \\
              \email{heye@kth.se, martin.monperrus@kth.csc.se}           
           \and
            Matias Martinez \at
            University of Valenciennes, France \\
            \email{matias.martinez@univ-valenciennes.fr}
}

\date{\vspace{-2cm}}

\maketitle

\begin{abstract}
In this paper, we do automatic correctness assessment for patches generated by program repair systems.
We consider the human-written patch as ground truth oracle and randomly generate tests based on it, a technique proposed by Shamshiri et al., called Random testing with Ground Truth (RGT) in this paper.
We build a curated dataset of 638 patches for Defects4J  generated by 14 state-of-the-art repair systems, we evaluate automated patch assessment on this dataset.
The results of this study are novel and significant:
First, we improve the state of the art performance of automatic patch assessment with RGT by 190\% by improving the oracle;
Second, we show that RGT is reliable enough to help scientists to do overfitting analysis when they evaluate program repair systems; 
Third, we improve the external validity of the program repair knowledge with 
the largest study ever.
\end{abstract}

\section{Introduction}
\label{sec:introduction}

Automatic program repair (APR) aims to reduce manual bug-fixing effort by providing automatically generated patches \citep{martinBibliography,TSE-repair-survey}. 
Most of the program repair techniques use test suites as a specification of the program, which is what we consider in this paper.
One of the key challenges of program repair is that test suites are generally too weak to fully specify the correct behavior of a program. Consequently, a generated patch passing all test cases may still be incorrect \citep{QiIssta15-overfitting}. 
Per the usual terminology, such an incorrect patch is said to be \textit{overfitting} if it passes all tests but is not able to generalize to other buggy input points not present in the test suite \citep{cure-worse-15}.
Previous research, e.g., \citet{Long-search-space}, was shown that automatic repair systems tend to produce more overfitting patches than correct patches.

Due to the overfitting problem, researchers cannot only rely on test suites to assess the capability of the new repair systems they invent.
Thus, a common practice in the program repair research community is to employ manual assessment for generated patches to assess their correctness. Analysts, typically authors of the papers, annotate the patches as `correct'  or `overfitting' \citep{martias2016defects4j} according to their analysis results. This assessment is typically done according to a human-written patch considered as a ground truth.
A patch is deemed as \textit{correct} if and only if:
1) it is identical to the human-written patch, 
or 
2) the analysts perceive it as semantically equivalent. 
Otherwise, a patch is deemed as \textit{overfitting}.

There are three major problems with manual patch assessment: difficulty, bias and scale.
First, in some cases, it is hard to understand the semantics of the program under repair. Without expertise on the code base, the analyst may simply be unable to assess their correctness \citep{martias2016defects4j,Yin-fse11}.
Second, the usual practice is that the analysts of patches are also authors of the program repair system being evaluated. Consequently, there may exist an inherent bias towards considering the generated patches as correct.
Third, it frequently happens that dozens of patches are generated for the same bug \citep{Le:overfitting,cardumen}. This makes the amount of required manual analysis quickly overpass what is doable in a reasonable amount of time.
For example, the recent work by \citet{RepairThemAll2019} resulted in more than \numprint{66596}  test-suite adequate patches for which it is impossible to manually assess their correctness.\footnote{The number of test-suite adequate patches are obtained from their experiment GitHub repository.}
To overcome difficulty, bias and scale in manual patch assessment, we need automated patch assessment \citep{XiongIdentifyingPC-ICSE18-patchsim,issta17-difftgen,le:reliability-patch-assess}.

Patch assessment is an indispensable task for the evaluation of a repair approach because it measures the \emph{effectiveness} of such an approach, which is subsequently reported in the academic literature. 
Inaccurate assessment affects both~
\begin{inparaenum}[\it a)]
\item  the progress of the research community because researchers discuss and compare program repair patches with different assessment criteria and methods,  and 
\item the adoption of program repair in practice because practitioners can have wrong expectations and can potentially underestimate the proportion of overfitting patches.
\end{inparaenum}
Overall, an effective automated patch assessment technique is significant for the program repair research.
First, it allows researchers to assess the correctness of program repair patches at scale, and this would enable them to compute the correctness labels of large datasets of patches. 
Second, having this large amount of annotated patches is a prerequisite to conducting wide analyses of the characteristics of overfitting patches, and to applying machine learning techniques that require a large amount of data.
Consequently, there is an important need for an automated patch assessment.

In this paper, we consider automated patch assessment given a ground truth reference patch written by humans, as done by  \citet{issta17-difftgen},  \citet{Le:overfitting} and  \citet{zhongxing-EMSE18}. Having a ground truth reference patch is in line with the manual assessment based on the human-written patch, and enables us to compare them. 
Notably, there exist other works such as those by  \citet{XiongIdentifyingPC-ICSE18-patchsim} and  \citet{Opad}  based on the opposite premise: the absence of a reference patch. 
\EMSErevision{We perform automated patch assessment technique using test generation: we generate tests based on the human-written patch, which encodes the correct program behavior considered as the oracles. If any automatically generated test fails on a machine patch, it means its behavior is different from the human-written patch's behavior and it is considered to be \textit{overfitting}. In this paper, we call this procedure RGT, standing for Random testing based on Ground Truth.}
Our implementation uses Evosuite \citep{evosuite} and Randoop \citep{randoop} as test generators and the collected 638 patches in our dataset are automatically assessed with \rgttotalnumber{} RGT tests generated (to our knowledge, the largest number of tests ever reported in this context).


Our large scale study enables us to identify 12 major findings that have important implications for future research in the field of automatic program repair: these findings and their implications are summarized in Table \ref{findings}. Thus, our work is novel as follows:
\begin{enumerate}
\item We show that 10 patches from previous research classified as correct by their respective authors are actually overfitting (as opposed to only one such a case in \citet{le:reliability-patch-assess});
\item We show that automated patch assessment is able to detect those manual errors, which is a key result for convincing the community to switch from manual patch assessment to automatic patch assessment;
\item We significantly increase the performance of automated patch assessment based on random testing with ground-truth (RGT): on the considered benchmark, the performance increase is 190\% higher than the state-of-the-art DiffTGen \citep{issta17-difftgen}; 
\item We measure the false positive rate of automated patch assessment which has never been done before;
\item Our study is at the largest scale ever, hence has a bigger external validity than the related work: we analyze 638 patches (versus 79 in \citet{issta17-difftgen} and 189 in \citep{le:reliability-patch-assess}) from 14 repairs systems (versus 6 considered in \citet{issta17-difftgen} and 8 considered in \citet{le:reliability-patch-assess}).
\end{enumerate}

Furthermore, we share with the program repair community:
1) a curated dataset of 638 patches generated from 14 program repair tools for the Defects4J benchmark, together with their correctness metadata. Those patches are given in canonical format so that they provide
a foundation for future program repair research. All the data presented in this paper are publicly-available\footnote{https://anonymous.4open.science/r/cffe573f-61ab-4d99-9e7c-dc769d657e75/}.
2) a curated dataset of \rgttotalnumber{} generated tests for Defects4J based on the ground truth human patch. This dataset is valuable for future automated patch assessment, as well as for sister fields such as fault localization and testing.

 \begin{table}[!h]
 \centering
 \footnotesize
 \renewcommand{\arraystretch}{1.0}
 \caption{ Our Major Findings and Their Implications Based on Our Study of 638 APR Patches and \rgttotalnumber{} Generated Tests for Automatic Patch Correctness Assessment. ``RGT'' Refers to the Patch Assessment Technique Based on Random Testing Which Introduced in \cite{ase15-tests} and Deepened in This Empirical Study.}
 \label{findings}
\hspace{-.5cm}\begin{tabular}{|p{0.5\linewidth}|p{0.5\linewidth}|}
\hline
\rowcolor{lightgray} \textbf{Findings on Manual Versus RGT Assessment} & \textbf{Implications} \\
\hline
\#F1 The misclassification of patches by manual assessment is a common problem.  Our experiment shows it happened for 6/14 repair systems manually assessed  in previous research. & 
(1) The research community of APR researchers needs better techniques for patch assessment to strengthen scientific validity. \\

\hline
\#F2 APR researchers confirm that the inputs sampled by random testing are valuable to assess patch correctness. & 

(2) It helps APR researchers to have concrete inputs to analyze patch correctness, suggesting more research about automatic identification of interesting input points (e.g. \citet{Shriver-inputs}). 
\\
\hline
\hline

\hline
\rowcolor{lightgray} \textbf{Findings on False Positive Ratio of RGT Assessment} & \textbf{Implications} \\
\hline
\#F3 RGT patch assessment sometimes suffers from false positives. In our experiment, the false positive rate of RGT is  6/257 (2.3\%). 
&
(3) This false positive rate is low, researchers can rely on RGT for providing better assessment results of their program repair contributions. \\
\hline

\#F4 RGT  causes false positive cases because the used test generation technique is not aware of preconditions or constraints on inputs. &
(4) Better support for preconditions in test generation would help to increase the reliability of RGT patch assessment. \\

\hline

\#F5 In our experiments, the RGT patch assessment yields three false positives because of optimization or  imperfection in the human-written patches.
& 
(5) We, as APR researchers, should not blindly consider the human-written patch as perfect ground truth, this impacts both manual assessment and automatic patch assessment.\\
\hline
\hline

\rowcolor{lightgray} \textbf{Findings on the Effectiveness of RGT Assessment} & \textbf{Implications} \\
\hline

\#F6 In our experiment, the RGT automatically identifies 274 / 381 (72\%) of patches claimed as overfitting by manual analysis. This is a significant improvement over \citep{le:reliability-patch-assess} in which fewer than 20\% of overfitting patches could be identified.
& (6) Our results suggest that the effectiveness of the RGT patch assessment was underestimated in \citet{le:reliability-patch-assess}. This calls for future research on this topic, with replication studies, in order to strengthen external validity.\\
\hline

\#F7 For RGT patch assessment, Evosuite outperforms Randoop in sampling inputs that differentiate program behaviors by 210\%, but considering these two techniques together can maximize the effectiveness of identifying overfitting patches. &
(7) Patch assessment techniques that involve automatic test generation can consider different techniques to maximize their effectiveness (e.g. PATCH-SIM \citep{XiongIdentifyingPC-ICSE18-patchsim}).   \\
\hline
\#F8 Behavioral differences identified with exception comparison is an important factor behind RGT's effectiveness. DiffTGen, which only considers assertion-based differences between output values, thus performs worse. & 
(8) Future overfitting detection techniques should consider both assertion and exception related behavioral differences.
\\

\hline
\#F9 We found flaky tests in both newly generated RGT tests and  previously generated RGT tests from previous research. &
(9) Flaky test detection is important to consider for RGT assessment. APR researchers who use RGT tests should give particular attention to identify flaky tests. \\
\hline
\hline

\rowcolor{lightgray} \textbf{Findings on RGT Time Cost } & \textbf{Implications} \\
\hline
\#F10 Over  87\%  of  the  time  cost  of  RGT  patch  assessment  is  spent  in
test case generation. &

(10) We encourage researchers to share the generated tests for behavioral assessment of APR patches.
This is a big time saver and this improves scientific reproducibility. \\

\hline
\#F11 Using previous generated RGT tests from \citep{ase15-tests} is able to identify 219/381 (57.5\%) of overfitting patches without paying any test generation time. &
(11) Future APR experiments on Defects4J can reuse previously generated RGT tests. When researchers assess patch correctness of APR patches with the same dataset of tests, the community has a fair and unbiased comparison of program repair effectiveness.
\\

\hline

\#F12 There is a trade-off between time spent in generating tests and effectiveness to discard overfitting patches.

&
(12) Our experiments provide practical configuration guidelines for future research and experiments using the RGT patch assessment technique. \\

\hline

\end{tabular}
\end{table}
\FloatBarrier

\section{Background}
\label{sec:bg}
This section provides our motivation through an example demonstrating the problem of manual patch assessment, as well as the background of the overfitting problem in program repair.

\subsection{Motivating Example}
\label{sec:motivating-example}
Manual patch assessment is an error-prone and subjective task, which could lead to various results depending on the knowledge and experience of the analysts.
\autoref{motivating-example} presents the human-written patch and the APR patch by Arja \citep{Yuan2017ARJAAR}, DeepRepair \citep{deeprepair}, and JGenProg \citep{martias2016defects4j} for  Chart-3 bug in Defects4J \citep{defect4j-dataset}.

\begin{listing}
\scriptsize
  
\caption{Motivating Example}
\noindent\begin{minipage}[b]{0.9\textwidth}
    \begin{lstlisting} [firstnumber=1056]
      TimeSeries copy =  (TimeSeries)super.clone();            
     <@\colorbox{lightgray}{+ copy.minY = Double.NaN;}@>                 
     <@\colorbox{lightgray}{+ copy.maxY = Double.NaN;}@>   
      copy.data = new java.util.ArrayList();
      if (this.data.size() > 0) { ...   
    \end{lstlisting}
    \subcaption{The human-written patch of bug Chart-3 in Defects4J}
    \label{motivating-example1}            
    \end{minipage}%
    \hfill
\begin{minipage}[b]{0.9\textwidth}
    \begin{lstlisting}[firstnumber=573]
      if (item == null) {
         throw new IllegalArgumentException("...");
      }
     <@\colorbox{lightgray}{+   findBoundsByIteration();}@>
      item = (TimeSeriesDataItem)item.clone();
    \end{lstlisting}
    \subcaption{The generated patch  by Arja, DeepRepair and  JGenProg}
    \label{motivating-example2}            
\end{minipage}
 \label{motivating-example}  
\end{listing}


Even though these three APR techniques generate the same patch for  bug Chart-3, however, their analysts hold different opinions about the correctness of the generated patch.
\autoref{tab:motiv-exmaple}  shows the assessment results for this APR patch from previous literature. Originally, the Arja analysts considered it as \textit{correct}, while it was deemed as \textit{overfitting} by the DeepRepair's analysts and \textit{unknown} by the JGenProg analysts.
\citet{le:reliability-patch-assess} employed 3 to 5 external software experts to evaluate the correctness of this patch and the result was \textit{overfitting}. 

We performed several discussions of the correctness of this patch with the original authors of DeepRepair and JGenProg via email. Eventually, they achieved consensus on the correctness of this patch and confirmed that this patch is actually a  \textit{correct} patch.  

 \begin{table*}[ht]
 \small
  \caption{Manual Analysis Result for Motivating Example}	
    \renewcommand{\arraystretch}{1.2}

   \label{tab:motiv-exmaple}
	\begin{tabular}{p{0.7\linewidth}lr}  
    \toprule
   Analysts &&Previous  Result \\
  \hline
    Arja \citep{Yuan2017ARJAAR} &&Correct \\
    DeepRepair  \citep{deeprepair} &&Overfitting\\
    JGenProg \citep{martias2016defects4j} &&Unknown\\
    3-5 Independent Annotators \citep{le:reliability-patch-assess}& &Overfitting\\
    \bottomrule
    \end{tabular}
     \end{table*}

The motivating example shows that analysts may hold different opinions of the correctness even on the same patch. 
If manual patch correctness assessment gives too many erroneous results, it is a significant threat to the validity of the evaluation of program repair research.
With unreliable correctness assessment, a technique A claimed as better than a technique B may actually be worse.
Ideally, we need a method that automatically and reliably assesses the correctness of program repair patches.

\subsection{Overfitting Patches}

Overfitting patches are those plausible patches that pass all developer provided tests, nevertheless, they fail to be a good general solution to the bug under consideration. As such, overfitting patches can fail on other held out tests \citep{cure-worse-15}. 
The essential reason behind the overfitting problem is that the test cases that are used for guiding patch generation are incomplete.

The overfitting problem has been reported both qualitatively and quantitatively in previous work \citep{cure-worse-15,QiIssta15-overfitting,Long-search-space,martias2016defects4j}. 
For example, in the context of Java code, 
\citet{zhongxing-EMSE18} studied the overfitting on Defects4J.
In the context of C code, \citet{Le:overfitting} measured that 73\% - 81\% of APR patches are overfitting considering two benchmarks, IntroClass and CodeFlaws.
\citet{QiIssta15-overfitting} conducted an empirical study on the correctness of three repair techniques. The three considered techniques have an overfitting rate ranging from 92\% to 98\%. Such a large percentage of overfitting patches motivates us to assess patch correctness in an automatic manner.

\subsection{Automated Patch Correctness Assessment}
\label{sec:kinds-of-assessment}
Typically, researchers employ the human-written patch as  ground truth to identify overfitting patches.
\citet{issta17-difftgen} propose DiffTGen to identify overfitting patches with tests generated by Evosuite \citep{evosuite}.
Those tests are meant to detect behavioral differences between a machine patch and a human-written patch. 
If any test case differentiates the output value between a machine patch and the corresponding human-written patch, the machine patch is assessed as overfitting.
DiffTGen has been further studied by \citet{le:reliability-patch-assess}, who have confirmed its potential.
Opad \citep{Opad} employs two test oracles (crash and memory-safety) to help APR techniques  filter out overfitting patches by enhancing existing test cases.
 \citet{XiongIdentifyingPC-ICSE18-patchsim} do not use a ground truth patch to determine the correctness of a machine patch.
They consider the similarity of test case execution traces to reason about overfitting.


\section{Experimental Methodology}
\label{sec:protocol}
In this section, we first present an overview of the RGT patch assessment (\ref{subsec:rgt}).
We then introduce seven categories of program behavioral differences for automated patch assessment (\ref{subsec:behavior-diff}) and present the workflow of the RGT assessment (\ref{subsec:rtg-approach}). 
After that, we present our research questions (RQs) to comprehensively evaluate the effectiveness and performance of RGT assessment (\ref{subsec:rqs}). Finally, 
we illustrate the methodology for each RQ in detail (\ref{subsec:protocols}).

 \begin{table*}[!t]
 \notsotiny
  \renewcommand{\arraystretch}{1.2}
  \caption{RGT Detects Seven Behavioral Differences}	
  \label{tab:rgt-oracles}
   \begin{tabular}{p{0.08\linewidth}p{0.255\linewidth}p{0.25\linewidth}p{0.32\linewidth}}
\toprule
 Differences & Ground-Truth Behavior  &Actual Behavior & Test Failure Diagnostic \\
\midrule
\multirow{2}{*}{$D_{assert}$}&\multirow{2}{*}{expect value $V1$} & \multirow{2}{*}{actual value $V2$ } & ComparisonFailure/AssertionError \\
&&&expected:  $V1$ but was: $V2$\\
$D_{exc1}$ & exception $E1$  & no exception  & Expecting exception: $E1$ \\
$D_{exc2}$&no exception   &  exception $E1$ & Exception $E1$ at  \\
$D_{exc\_type}$& exception   $E1$ &  exception $E2$ & Expected exception of type $E1$ \\
\multirow{2}{*}{$D_{exc\_loc}$}&\multirow{2}{*}{exception $E1$ by function \tiny{$A$} }& \multirow{2}{*}{exception $E1$ by function \tiny{$B$} } & Expected exception $A.E1$ \\
&&& but was $B.E1$\\
$D_{timeout}$&execution within timeout \tiny{$T$} & execution out of timeout \tiny{$T$} & Test timed out after \tiny{$T$} milliseconds\\

$D_{error}$& no error   & error &  Other failures \\
\bottomrule
\end{tabular}	
\end{table*}

\subsection{An Overview of RGT Patch Assessment }
\label{subsec:rgt}
The goal of the RGT patch assessment is to automatically assess the correctness of APR patches.
It is based on 1) a ground truth patch and 2) a random test generator.
The intuition is that random tests would differentiate the behaviors between a ground truth patch and an APR patch.
\EMSErevisionMinor{In our work, we consider the human-written patched program as system under test (SUT) because it encodes human knowledge and domain-specific expertise for fixing the bug. On the contrary, we do not consider the buggy program as SUT because it is a bad oracle (see the oracle problem discussed in \cite{oracleproblem}):
considering the buggy program as SUT would encode the incorrect behaviors in the generated tests, which would then mislead patch correctness assessment.
}

With regard to test generation,  we consider  typical regression test generation techniques \citep{randoop,evosuite} for randomly sampling regression oracles based on a ground truth program. 
In other words, these automatic test case generation techniques use the current behavior of the program itself as an oracle \citep{regressionSurvey, xietao}. 
\EMSErevisionMinor{In our experiment, we use the regression mode of Evosuite \citep{evosuite} and Randoop \citep{randoop}, which consists of creating assertions in the generated tests.
}
Consequently, a ``RGT test'' in this paper refers to a test generated based on a ground truth patch, containing oracle that encodes runtime behaviors of a ground truth program (\EMSErevisionMinor{i.e., human-written patched program}).

RGT patch assessment takes RGT tests and an APR patched program as inputs and outputs the number of test failures that witness a behavioral difference.
RGT patch assessment establishes a direct connection between the outputs of random tests and overfitting classification:
if any behavioral difference exists between an APR patch and a ground truth patch, such APR patch is assessed as overfitting. 
More specifically, if a ground truth patch passes all RGT tests 
but an APR patch fails on any of them, this APR patch is assessed as overfitting. 
While RGT patch assessment is a known technique, it has not been studied at a large scale.

\subsection{Categorization of Behavioral Differences}
\label{subsec:behavior-diff}
Based on our experiment of executing \rgttotalnumber{} RGT tests on 638 patches, we empirically define seven program behavioral differences that could be revealed by RGT tests. They are summarized in \autoref{tab:rgt-oracles}. 
The first column gives the identifier of differences between the ground truth program behavior (shown in the second column) and the actual patched program behavior (shown in the third column). 
In the fourth column, we give the test failure diagnostic that is used for mapping each category. 
In our study, we use regex patterns to match test failure diagnostics that enable us to automatically classify the behavioral difference categories.

Now we explain them as follows:


\emph{$D_{assert}$}: Given the same input, the expected output value from the ground truth program is different from the actual output value from the patched program. In this case, a difference in value comparison reveals an overfitting patch.

\emph{$D_{exc1}$}: Given the same input, an exception is thrown when executed on the ground truth program but the patched program does not throw any exception when executed with such input. The expected behavior is an exception in this case.

\emph{$D_{exc2}$}: Given the same input, no exception is thrown when executed on the ground truth program but at least one exception is thrown when executed on the patched program. The expected behavior is no exception in this case. 

\emph{$D_{exc\_type}$}: Given the same input, an exception \emph{E1} is thrown when executed on the ground truth program but a different exception \emph{E2} is thrown when executed on the patched program.  The expected behavior is the exception \emph{E1} in this case. 

\emph{$D_{exec\_loc}$}: Given the same input, an exception \emph{E1} is thrown by the function \emph{A} when executed on the ground truth program but the same exception \emph{E1} is thrown by another function \emph{B} when executed on the patched program. In this case, we consider the same exception produced by different functions as behavioral differences.

\emph{$D_{timeout}$}: Given the same input and a large enough timeout configuration value $T$, the ground truth program executed within a considered timeout but the execution of the patched program causes a timeout.

\emph{$D_{error}$}: 
Given the same input, no error is caused when executed on the ground truth program but an error is caused when executed on the patched program. $D_{error}$ indicates an unexpected error while test execution, instead of a test failure. 
The cause of a test error can be various. In this study, we consider failing tests not mapped in the aforementioned six categories belong to $D_{error}$.

\subsection{The RGT Algorithm}
\label{subsec:rtg-approach}
The RGT algorithm has been proposed by  \citep{ase15-tests}. It consists of using generated tests to identify a behavioral difference. We use it in the context of a patch assessment process for program repair.
Algorithm 1 presents the RGT algorithm. 
RGT takes as input a machine patch set $P$, a ground truth patch set $G$, and the automatically generated RGT test set $T$.
As a result, RGT outputs for each machine patch from $P$ two diagnoses: a) a $label$, which is either correct or overfitting, b) a list of behavioral differences.
The assessment process mainly consists of two procedures that we discuss now: sanity check for $T$ and automatic assessment for $P$.

\emph{Sanity Check:} 
We first perform a sanity check for RGT tests in  $T$ in order to detect and remove flaky tests, those generated tests that have non-deterministic behaviors.  
For each human-written patched program $p_{hi}$ from $G$, 
we execute the corresponding RGT tests $T_{i}$ against $p_{hi}$. 
If any test in $T_{i}$ yields a failure against $p_{hi}$, we add it into a flaky test set $FLAKY_{i}$ (line 7). 
If  $FLAKY_{i}$ captures any flaky test, we then remove all tests in $FLAKY_{i}$ from  $T_{i}$ (line 8). 
We conduct this procedure consecutively $n$ times to maximize the likelihood of detecting flaky tests ($n$ is the $cnt$ variable at line 4, it is set to 3).

\emph{Assessment:}
For the considered patch set $P$ and RGT test set $T$, after the sanity check (line 11),  we execute all tests from $T$  against each machine patch in $P$.
If any generated test yields a failure against a machine patch $p_{mi}$, it is recorded in the 
failing test set $FT_{i}$ (line 13), signaling a behavioral difference.
If the $FT_{i}$ captures the failing test, the correctness label of such $p_{mi}$ is set to \textit{overfitting}, otherwise \textit{correct}.
Regarding the patches assessed as \textit{overfitting}, for each failing test,  we analyze the failure and add one of the seven categories of behavioral differences in the set $D_{p_{mi}}$ according to their failure diagnostic (line 18). 
As a result, RGT outputs the correctness label and a set of behavioral differences for each machine patch.

\begin{algorithm}[H]
\textbf{Input:} \textbf{(1)} the machine patch set P=\{$p_{m1}$...$p_{mn}$\}, where ~$p_{mi}$ is a machine patched program for bug $i$; \textbf{(2)} a ground truth patch set G=\{ $p_{h1}$... $p_{hk}$\},  where ~ $p_{hi}$ is a ground truth patch for bug $i$; \textbf{(3)} RGT test set T = \{$T_{1}$...$T_{k}$\}, where~ $T_{i}$ is a set of tests generated for bug $i$.

\textbf{Output:}  the correctness label:correct/overfitting; a list of behavioral differences
\caption{RGT Patch Assessment }
\begin{algorithmic}[1]

\Procedure{SanityCheck}{$G,T$} 
\For{$p_{hi}$ in G}
\For{$T_{i}$ in T}
\State $cnt \gets 3$
\While{$cnt > 0$}
\State $cnt \gets cnt-1$
\State $FLAKY_{i} \gets  runTests(p_{hi}, T_{i})$ 
\State $T_{i}$ = $T_{i} - FLAKY_{i}$
\EndWhile
\EndFor
\EndFor
\Return $T$
\EndProcedure
\Procedure{Assessment} {$G,P,T$}
\State $A_r \gets \emptyset$
\State $T \gets SanityCheck (G,T)$
\For{$P_{mi}$ in $P$}
\State $FT_{i} \gets runTests(p_{mi}, T_{i})$
\If{\texttt{$FT_{i} \neq \emptyset $}} 
\State $ label_{P_{mi}} \gets overfitting $
\State $D_{p_{mi}} \gets  \emptyset$
\For{$t_{i}$ in $FT_i$}
\State $D_{p_{mi}} \gets D_{p_{mi}} \cup   t_{i}$
\EndFor
\State $A_r \gets  A_r  	\cup  \langle P_{mi}, label, D_{p_{mi}}  \rangle$
\Else{}
\State $ label_{P_{mi}} \gets correct $
\State $A_r \gets  A_r  	\cup  \langle P_{mi}, label, null  \rangle$
\EndIf
\EndFor
\Return $A_r$
\EndProcedure
\end{algorithmic}
\end{algorithm}

\subsection{Research Questions}
\label{subsec:rqs}
We intend to comprehensively evaluate the effectiveness of the RGT patch assessment.
For this, we investigate the following RQs:

\begin{itemize}
\item RQ1: \RQone{} This is a key to see whether RGT patch assessment is better than manual patch assessment or rather complementary. We also ask researchers from the program repair community about the misclassification cases.

\item RQ2: \RQtwo{} There are a number of pitfalls with  RGT patch assessment which have never been studied in depth.

\item RQ3: \RQthree{}
\item RQ4: \RQfour{}
Also, we study whether we could reuse tests generated in previous research projects to speed up the patch assessment process.
\item RQ5: \RQfive
\end{itemize}

\subsection{Protocols}
\label{subsec:protocols}

\textbf{RQ1 \& RQ2.} 
We first collect a set of APR patches for Defects4J, that were claimed as correct by their respective authors.
This set of patches is denoted as $D_{correct}$.
Next, we execute RGT tests over all $D_{correct}$ patches and we report the number of patches that make at least one RGT test fail.
This case means that the RGT patch assessment contradicts the manual analysis previously done by APR researchers.
Then, we manually investigate the cases where a patch is classified as overfitting by RGT assessment. 
This manual analysis aims at separating true positives from false positives.
Our manual analysis is performed as follows:
we first manually compare those patches flagged as overfitting with the human-written patches and give the patch one of four labels: no or partially fix, a new bug introduced, semantically equivalent, and identical. 
In addition, we analyze the RGT test failures to determine (1) whether the observed behavioral difference is indeed triggered by an incorrect patch and (2) whether the sampled test data violates any program pre-condition.  
All results are discussed among at least two authors.
The RGT assessment is considered as a true positive if:
a) our manual analysis perceives the patch is `no or partially fix' or `a new bug introduced', 
b) the observed behavioral difference is related to the bug, 
and
c) the sampled test data does not violate any program pre-condition.
Last, for those true positive cases, we send our RGT assessment results and failing RGT tests to the original authors of the patch and ask them for feedback.
In particular, we explore to what extent they agree with the RGT assessment results. 
For RQ2, we record the number of the false positive cases by RGT assessment. This enables us to estimate a false positive rate of RGT assessment.

\textbf{RQ3.} RQ3 focuses on the effectiveness of RGT assessment in identifying overfitting patches.
We first collect a set of APR patches for Defects4J, that were manually assessed as overfitting by the corresponding researchers. This set of patches is denoted as $D_{overfitting}$.
We execute RGT tests over the whole $D_{overfitting}$ patches and record test failures.
A test failure means that RGT succeeds in identifying a patch as overfitting, that RGT agrees with the manual analysis by researchers.
Next, we also execute the state-of-the-art overfitting patch detection technique DiffTGen over the same dataset. We execute DiffTGen by the default mode which calls EvoSuite in 30 trials with the searching timeout being 60 seconds for each trial.
We do not execute Opad \citep{Opad} and PATCH-SIM \citep{XiongIdentifyingPC-ICSE18-patchsim} on this dataset for the following reasons: 
Opad is based on memory safety analysis in C which is not relevant in the context of the memory safe language Java.
PATCH-SIM is not considered in our study for two reasons:
(1) PATCH-SIM  targets APR users who do not have any ground truth patch available. On the contrary, RGT targets APR researchers who have a ground truth patch at hand.
As reported in \citet{XiongIdentifyingPC-ICSE18-patchsim},  PATCH-SIM has  a false positive rate of 8.25\% for assessing human-written patches, we aim at having a lower false positive rate.

\textbf{RQ4.} We estimate the performance of RGT from a time cost perspective. 
We measure the time cost of RGT in three dimensions: the time cost of test case generation, the time cost of sanity checking and the time cost of executing the test cases over the APR patches.
Those three durations are respectively denoted $TCGen$, $SC$, and $EXEC$.
 Next, we collect RGT tests from previous research. 
Last, we execute previously generated RGT tests over both $D_{correct}$ and $D_{overfitting}$ in order to compare both $SC$ and $EXEC$.   We evaluate the effectiveness of previously generated RGT tests by comparing them with the new generated RGT tests.

\textbf{RQ5.}  RQ5 investigates the trade-off between the number of RGT test generation costs and the effectiveness of overfitting patch classification.
We conduct our experiment of executing 30 runs of RGT tests on  $D_{overfitting}$. 
First, we record the number of overfitting patches individually identified by each test generation. 
Next, to account for randomness, we analyze 1000 random groups and each of which is with a random sequence of 30 test generations. 
Last,
we analyze the number of test generations on average and their effectiveness of overfitting patch identification.

\begin{table*}  \scriptsize
  \caption{Dataset of Collected Defects4J Patches}	
  \label{drr-patches}
  \renewcommand{\arraystretch}{1.2}
   \begin{tabular}{p{0.16\linewidth} p{0.17\linewidth} rrrrr|c}
    \toprule
    Dataset  &  APR Tool  &  Chart  &  Closure  &  Lang  &  Math  &  Time  &   Total\\
    \toprule
 \multirow{13}{*}{\scriptsize \textbf{$D_{correct}$}}  &  ACS  &  2  &  0  &  3  &  12  &  1  &  18\\
  &  Arja   &  3  &  0  &  4  &  10  &  1  &  18\\
  &  CapGen   &  5  &  0  &  9  &  14  &  0  &  28\\
  &  DeepRepair   &  0  &  0  &  4  &  1  &  0  &  5\\ 
  &  Elixir   &  4  &  0  &  8  &  12  &  2  &  26\\    
  &  HDRepair   &  0  &  0  &  1  &  4  &  1  &  6\\
  &  Jaid   &  8  &  9  &  14  &  11  &  0  &  42\\  
  &  JGenProg2015   &  0  &  0  &  0  &  5  &  0  &  5\\   
  &  Nopol2015   &  1  &  0  & 3  &  1  &  0  &  5\\
  &SequenceR & 3 & 4 & 2 & 8 &0 & 17\\
  &  SimFix   &  4  &  6  &  9  &  14  &  1  &  34\\
  &  SketchFix   &  6  &  2  &  2  &  6  &  0  &  16\\
  &  SOFix   &  5  &  0  &  3  &  13  &  1  &  22\\
  &  ssfix   &  2  &  1  &  5  &  7  &  0  &  15\\
\hline
 \multicolumn{2}{c}{Sum for $D_{correct}$}  &  43  &  22  &  67  &  118  &  7  & 257\\
\hline
   
\multirow{13}{*}{\scriptsize \textbf{$D_{overfitting}$}}  &  ACS   &  0  &  0  &  1  &  4  &  0  &  5\\
  &  Arja   &  30  &  0  &  54  &  73  &  15  &  172\\
  &  CapGen   &  0  &  0  &  14  &  24  &  0  &  38 \\
  &  DeepRepair   &  4  &  0  &  1  &  4  &  0  &  9\\ 
  &  Elixir   &  3  &  0  &  4  &  7  &  1  &  15\\
  &  HDRepair  &  0  &  0  &  0  &  3  &  0  &  3\\
  &  Jaid   &  8  &  4  &  10  &  17  &  0  &  39\\  
  &  JGenProg2015   &  3  &  0  &  0  &  2  &  1  &  6\\     
  &  Nopol2015    &  0  &  0  &  2  &  3  &  1  &  6\\
  & SequenceR &  3 & 32  &   1 &  20 & 0           & 56 \\ 
  &  SimFix   &  0  &  0  &  3  &  9  &  0  &  12\\   
  &  SketchFix   &  2  &  0  &  2  &  5  &  0  &  9\\   
  &  SOFix   &  0  &  0  &  0  &  2  &  0  &  2\\
  &  ssfix   &  1  &  1  &  1  &  6  &  0  &  9\\  
\midrule
\multicolumn{2}{c}{Sum for $D_{overfitting}$}  &  54  &  37  &  93  &  179  &  18  &  381\\ \hline
\multicolumn{2}{c}{Sum for all}  &  97  &  59  &  160 & 297  & 25  &  638\\ \hline
 
\end{tabular}	
\end{table*}

\subsection{Curated Patch Dataset}
\label{intro-patch-collection}
\emph{Fourteen repair systems.}
APR patches for Defects4J form the essential data for our experiment.
The criteria of repair systems considered in this study are that they were previously evaluated on the Defects4J \citep{defect4j-dataset} benchmark.

We carefully collect APR patches that are publicly available. 
We perform this by browsing the repositories / appendices / replication packages of the corresponding research papers or by asking the authors directly. 
As a result, we build our dataset $D_{correct}$ and $D_{overfitting}$ from following 14 APR systems: 
ACS \citep{Xiong-ACS-ICSE17}; Arja \citep{Yuan2017ARJAAR}; CapGen  \citep{capgen-ICSE18}; DeepRepair  \citep{deeprepair}; Elixir \citep{elixir}; HDRepair  \citep{hdrepair}; Jaid \citep{jaid}; JGenProg \citep{martias2016defects4j}; Nopol \citep{martias2016defects4j};
SimFix  \citep{Simfix:2018}; SketchFix  \citep{sketchfix}; SOFix \citep{sofix}; ssFix \citep{ssFix}; SequenceR \citep{sequencer}.

\textit{Patch  Canonization and Verification.} 
In order to fully automate RGT patch assessment, we need to have all patches in the same canonical format. Otherwise, applying a patch may fail for spurious reasons. 
To do so, we manually convert the collected patches from their initial formats, such as XML, plain log file, patched program and etc., into a unified DIFF format. 
After unifying the patch format, we carefully name the patch files according to a systematic naming convention: \textbf{\small $<$PatchNo$>$-$<$ProjectID$>$-$<$BugID$>$-$<$ToolID$>$.patch}. For instance, \textit{patch1-Lang-24-ACS.patch} refers to the first patch generated by ACS to repair the bug from the Lang project identified as 24 in Defects4J.

\textit{Sanity  Check.}
Some shared patches may not be plausible per the definition of test-suite based program repair (passing all test cases).
We conduct a rigorous sanity check to  keep applicable and plausible patches. 
\textit{Applicable} means that a patch can be  applied successfully for the considered Defects4J version\footnote{Version 1.2: commit at 486e2b49d806cdd3288a64ee3c10b3a25632e991}. 
\textit{Plausible}  means that a patch is test-suite adequate, we check this property by executing the human-written test cases originally provided by Defects4J.
Eventually, we discard all patches that are not applicable or not plausible.

\subsection{Curated Dataset of Ground Truth based Random Tests}

Now we present our curated dataset of RGT tests generated based on ground truth patched programs. We consider both the previously generated RGT tests and newly generated RGT tests in our study.

\subsubsection{Previously Generated RGT Tests}
We search and obtain existing generated test cases for Defects4J from previous research.
\begin{itemize}
\item 
$Evosuite_{ASE15}$ : tests generated by Evosuite from ASE'15 paper \citep{ase15-tests}; 
\item $Randoop_{ASE15}$: tests generated by  Randoop from ASE'15 paper  \citep{ase15-tests}; 
\item $Evosuite_{EMSE18}$: tests generated by Evosuite from EMSE'18 paper \citep{zhongxing-EMSE18}.
\end{itemize}

$Evosuite_{ASE15}$ and $Randoop_{ASE15}$ were generated for 357 Defects4J bugs and each of them with 10 runs of test generation (with 10 seeds).  $Evosuite_{EMSE18}$ were generated for 42 bugs with 30 runs of test generation (with 30 seeds).

\subsubsection{New Generated RGT Tests} 

In this paper, we decided to generate new RGT tests for two main reasons.
First, we execute 30 runs of Evosuite \citep{evosuite} and Randoop \citep{randoop}, using a different random seed value on each, with the goal of generating new test cases (not generated by the 10 executions from \citep{ase15-tests}) that potentially detect behavioral differences.
They are respectively denoted as $RGT_{Evosuite2019}$ and $RGT_{Randoop2019}$.
By using 20 additional executions with new seeds, the new test cases sample other parts of the input space. 
Second, the test dataset from $Evosuite_{EMSE18}$ partially covers the Defects4J bugs (42 in total).

\emph{Parameters} 
We run both Evosuite and Randoop
on the ground truth program with 30 different seeds according to \citep{randomness-guide}, we consider 30 runs  in line with previous studies \citet{le:reliability-patch-assess,zhongxing-EMSE18,issta17-difftgen}.
We configure a timeout of 300 seconds and a search budget of 100 seconds for each test execution. 
In this paper, contrary to \citep{ase15-tests}, we did not consider the test generation tool AgitarOne because it requires a licensed infrastructure and it requires manual effort to generate and analyze a test suite.
We consider the branch coverage to guide RGT test generation which has been reported as the most effective coverage metric for fault detection compared with the other seven coverage metrics  \citep{test-ibf20}.

\subsubsection{Sanity Check}

Per the aforementioned RGT approach in \autoref{subsec:rtg-approach}, we conduct the sanity check for both previous generated RGT tests and newly generated RGT tests.
We execute each RGT test consecutively three times over the ground truth program. If any test yields a failure against the ground truth program, we discard it until all RGT tests pass three consecutive sanity checks. By doing so, we obtain a set of stable RGT tests for assessing patch correctness.


\section{Experimental Results}
\label{sec:result}

We now present our experimental results. We first look at the dataset and RGT tests we have collected.

\subsection{Patches}

We have collected a total of 638 patches  from 14 APR systems.
All pass the sanity checks described in \autoref{intro-patch-collection}. 
\autoref{drr-patches} presents this dataset of patches for Defects4J.
The first column specifies the dataset category and the second column gives the name of the automatic repair system.  The number of patches collected per project of Defects4J is given in the third to the seventh columns and they are summed at the last column.
They are 257 patches previously claimed as correct, which form $D_{correct}$.
There are 381 patches that were considered as overfitting by manual analysis in previous research, they form $D_{overfitting}$.
We found 160/257 patches from $D_{correct}$  are syntactically equivalent to the human-written patches: the exact same code modulo formatting, and comments.
The remaining 97/257  patches are semantically equivalent to human-written patches. Overall, the 638 patches cover 117/357 different bugs of Defects4J.\footnote{version 1.2: commit at 486e2b49d806cdd3288a64ee3c10b3a25632e991}
To our knowledge, this is the largest ever APR patch dataset with manual analysis labels by the researchers. The most related dataset is from \citep{issta17-difftgen} containing 89 patches from 4 repair tools and the one from  \citep{XiongIdentifyingPC-ICSE18-patchsim} containing 139 patches from 5 repair tools. Our dataset is 4 times larger than the latter.

\subsection{Tests} 
Evosuite and Randoop have been invoked 30 times with random seeds for each of the 117 bugs covered by the patch dataset. In total, they have been separately invoked for $117 \mbox~ bugs \times 30 \mbox ~ seeds=3510~ runs$. 

\subsubsection{Coverage} 

\begin{figure}
\centering
\caption{Code Coverage Distribution of RGT Tests}
\includegraphics[width=0.8\textwidth]{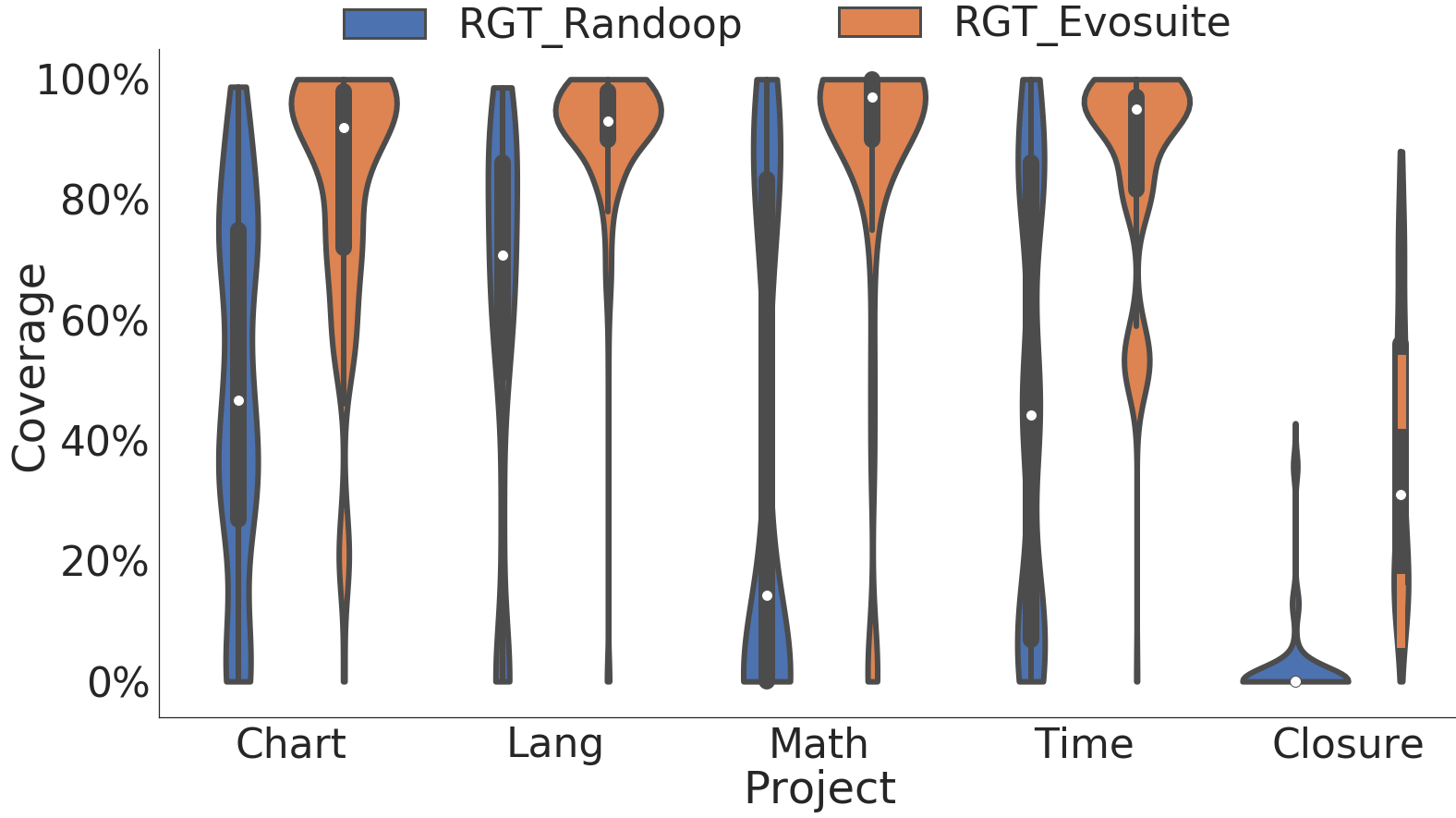}  
\label{fig:coverage}
\end{figure}

To better understand the generated RGT tests, we compute their coverage over the buggy classes. 
\autoref{fig:coverage} illustrates the code coverage distribution on the buggy classes by 3510 generated test suites in five Defects4J projects. 
We use Jacoco\footnote{https://www.jacoco.org, accessed February 2020} to measure the branch coverage on the buggy classes.
The orange legend shows the code coverage distribution by $RGT_{Evosuite2019}$ test suites while the blue one represents the coverage of $RGT_{Randoop2019}$ test suites. 
For example, in the Chart project, the code coverage ratios of $RGT_{Evosuite2019}$ are mostly over 80\% while the coverage of $RGT_{Randoop2019}$  is uniformly distributed between 0\% and 100\%. Therefore, the code coverage by $RGT_{Evosuite2019}$  is considered higher than the $RGT_{Randoop2019}$ .


Overall five projects, we observe that the code coverage by the $RGT_{Evosuite2019}$ is higher than the code coverage by the $RGT_{Randoop2019}$ test suites. 
For Chart, Lang, Math and Time projects, $RGT_{Evosuite2019}$ test suites achieve high code coverage on the buggy classes: the 90\% percentile is higher than 80\%. 
On the contrary,  the code coverage by $RGT_{Randoop2019}$  is clearly lower. 
The reasons are as follows:
First, $RGT_{Evosuite2019}$ suffers from fewer test generation failures: among 3510 random test suite generations, Evosuite fails to produce RGT tests in 31 runs while  Randoop fails in 1080 runs, which lead to a respectively a 0.9\% and  a 30.8\% failure rate.
Second, Evosuite applies a genetic algorithm in order to
evolve test cases that maximize code coverage, which has been consistently shown to be better than Randoop \citep{ase15-tests,automatic-test-competition}.


Notably, the code coverage on the Closure project is significantly lower than for the other four projects, both for $RGT_{Evosuite2019}$ and $RGT_{Randoop2019}$ test suites.
We found two reasons that explain that:
1) the Closure project requires test data with a complex data structure, which is a known hard challenge for automatic test generators;
2) the Closure project has a majority of private methods, which are not well handled by the considered test generation tools.

\subsubsection{Flaky Tests} 
We discard 2.2\% and 2.4\% flaky tests from $RGT_{Evosuite2019}$  and $RGT_{Randoop2019}$  respectively with a strict sanity check. 
As a result, we have obtained a total of \numprint{4477707} stable RGT tests: \numprint{199871} by $RGT_{Evosuite2019}$ and  \numprint{4277836} by $RGT_{Randoop2019}$.

We also collect  RGT tests generated by previous research, they are
\numprint{15136567} tests:  \aseevogeneratedvalidtests in  $RGT_{EvosuiteASE15}$ \citep{ase15-tests},  
\aserandoopgeneratedvalidtests in $RGT_{RandoopASE15}$ \citep{ase15-tests}, 
and \emseevogeneratedvalidtests{} in $RGT_{EvosuiteEMSE18}$ \citep{zhongxing-EMSE18}. By conducting a sanity check of those tests, we discard 2.7\%, 4.7\% and 1.1\% flaky tests. Compared with the newly generated RGT tests, more flaky tests exist in previous generated tests due to external factors such as version, date and time \citep{ase15-tests} (\#F9).
To our knowledge, this is the largest ever curated dataset of generated tests for Defects4J.

\subsection{Result of RQ1: RGT Patch Assessment Contradicts Previously Done Manual Analysis }
\label{rq1-sec}

 \begin{table*}[t]
  \centering
  \scriptsize
   \renewcommand{\arraystretch}{1.4}
  \caption{Misclassified  Patches Found by RGT. The Original Authors Agreed with the Analysis Error.}
   \label{rq1-result}
	\begin{tabular}{p{0.3\linewidth} ccl p{0.16\linewidth} }
    \toprule
&\multicolumn{2}{c}{RGT Tests}&&\\
\cline{2-3}
Patch Name&$Evos2019$&$Rand2019$&Category&Consensus\\
 \hline
 patch1-Lang-35-ACS&12&140&$D_{exc2}$&confirmed\\   
\hline
patch1-Lang-43-CapGen& 10 & 0 & $D_{error}$&confirmed  \\
\hline
patch2-Lang-43-CapGen&10 & 0 & $D_{error}$ &confirmed  \\
\hline
patch2-Lang-51-Jaid&43&0&$D_{assert}$&confirmed\\
\hline 
patch1-Lang-27-SimFix&32&0&$D_{exc1}$&confirmed \\
\hline
patch1-Lang-41-SimFix&124&0&$D_{assert}$&confirmed \\
\hline
patch1-Chart-5-Nopol2015&1&266&$D_{exc2}$&confirmed\\
\hline
patch1-Math-50-Nopol2015&2&0 &$D_{exc1}$  &confirmed\\
\hline
patch1-Lang-58-Nopol2015&21&0&$D_{assert}$& confirmed \\

\hline
\multirow{2}{*}{patch1-Math-73-JGenProg2015}& \multirow{2}{*}{49}&\multirow{2}{*}{0}&  $D_{exc1}$  & \multirow{2}{*}{confirmed}\\
&&&$D_{assert}$&\\
    \hline
    Sum  &10 &2 & - & 10 confirmed\\
    \hline
    \end{tabular}
    \end{table*}

We have executed 30 runs of RGT tests over 257 patches from $D_{correct}$. 
For the 160 patches syntactically equivalent to the ground truth patches, the results are consistent: no RGT test fails.
For the remaining 97 patches, the assessment of 16 patches contradicts previously reported manual analysis (at least one RGT test fails on the patch considered as correct in previous research).
This makes 10/16 true positive cases while the 6/16 are false positives according to our manual analysis. 
Due to the potential risk of false negatives with RGT tests, we also manually analyze the remaining 81 semantically equivalent patches which do not make any RGT test fail, the result is discussed in \autoref{dis-manual-assess}.

The ten true positive cases are presented in \autoref{rq1-result}.
The first column gives the patch name, with the number of the failing test by each RGT category in the second and third columns. The fourth column shows the category of behavioral difference defined in \autoref{tab:rgt-oracles}.
The last column gives the result of the conversation we had with the original authors about the actual correctness of the patch.
For instance, the misclassified patch of \textit{patch1-Lang-35-ACS} is identified as overfitting by 10 tests from $RGT_{Evosuite2019}$ and it is exposed by behavioral difference category $D_{exc2}$ of non-semantically behavior: no exception thrown from a ground truth program but exceptions caused in a patched program execution.
This result has been confirmed by the original authors.

$RGT_{Evosuite2019}$ and $RGT_{Randoop2019}$ identify 10 and 2 misclassified patches individually. This means that Evosuite is better than Randoop on this task. 
Now we look at the behavioral differences of those 10 misclassified patches which are exposed by four categories of behavioral differences. This shows the diversity of behavioral differences is important for RGT assessment.

Notably, the 10 misclassified patches are from 6/14 repair systems, which shows the misclassification in manual patch assessment is a common problem (\#F1). 
This shows the limitation of the manual analysis of patch correctness.
The 10.3\% (10/97) previously claimed correct semantically equivalent patches were overfitting, which shows that manual assessment of semantical APR patches is hard and error-prone. 
A previous research \citep{Shangwen19ESEM} reported over a quarter of correct APR patches are actually semantic patches, and this warns us should pay careful attention in assessing their correctness.
All patches have been confirmed as misclassified by the original authors. 
Five researchers gave us feedback that the inputs sampled by the RGT technique were under-considered or missed in their previous manual assessment. The RGT assessment samples corner cases of inputs that  assist researchers in manual assessment (\#F2).

We now present a  case study to illustrate how those patches are assessed by RGT tests.

\begin{listing}

\caption{The Case Study of Two Patches were Misclassified}
\label{case-study-lang-43}
\noindent\begin{minipage}[b]{0.9\textwidth}
    \begin{lstlisting} [firstnumber=419] 
       int start = pos.getIndex();
       char[] c = pattern.toCharArray();
       if(escapingOn && c[start] == QUOTE){
     <@\colorbox{lightgray}{+      next(pos); }  @>
    \end{lstlisting}
    \subcaption{The human-written patch for Lang-43}
    \label{patch-Lang-43-ground-truth}            
    \end{minipage}%
    \hfill
\begin{minipage}[b]{0.9\textwidth}
    \begin{lstlisting}[firstnumber=419] 
       int start = pos.getIndex();
       char[] c = pattern.toCharArray();
     <@\colorbox{lightgray}{+      next(pos);   }@>
       if(escapingOn && c[start] == QUOTE){
  
    \end{lstlisting}
    \subcaption{The generated patch of patch1-Lang-43-CapGen}
    \label{patch-Lang-43-1}            
\end{minipage}
   \hfill
   \begin{minipage}[b]{0.9\textwidth}
    \begin{lstlisting}[firstnumber=419] 
       int start = pos.getIndex();
     <@\colorbox{lightgray}{+      next(pos);   }@>
       char[] c = pattern.toCharArray();
       if(escapingOn && c[start] == QUOTE){
    \end{lstlisting}
    \subcaption{The generated patch of patch2-Lang-43-CapGen}
    \label{patch-Lang-43-2}            
\end{minipage}

\end{listing}

\emph{Case study of Lang-43.}
The CapGen repair tool generates three patches for bug Lang-43.
Those three patches are all composed of a single inserted statement \texttt{next(pos)} but the insertion happens at three different positions in the program.
Among them, there is one patch that is identical to the ground truth patch (\autoref{patch-Lang-43-ground-truth}). It inserts the statement in an if-block.
Patches \textit{ patch1-Lang-43-Capgen} (\autoref{patch-Lang-43-1}) and \textit{patch2-Lang-43-Capgen} (\autoref{patch-Lang-43-2}) insert the correct statement but at different locations, respectively 1 line and 2 lines before the correct position from the ground truth patch. 
Both patches are classified as overfitting by RGT, because 10 sampled inputs result in a heap space error. With the same inputs, the ground truth patch performs without exception, 
this corresponds to the category $D_{error}$ in \autoref{tab:rgt-oracles}.
The original authors have confirmed the misclassification of these two patches. This case study illustrates the difficulty of APR patch assessment: it is unlikely to detect a heap memory error by only reading over the source code of the patch.

\begin{mdframed}
Answer to RQ1: 
Among the 257 APR patches claimed as correct in previous work, 160 are syntactically identical to the human written patch, and 97 were assessed as semantically equivalent to the human written patch. Using automated patch assessment with RGT, we find that 10/97 (10\%) are actually overfitting.  All 10 patches have been confirmed as actually overfitting by their original authors. This clearly shows that manual analysis of the correctness of APR patches is hard and error-prone. 
The most closely related experiment is the one performed by \citep{le:reliability-patch-assess}, which is based on 45 claimed correct patches (as opposed to 257) and where one single patch is identified as misclassification (as opposed to 10). 
Our experiment significantly improves the external validity of this scientific finding as it is performed on a five times larger dataset.
\end{mdframed}

\subsection{Result of RQ2: False Positives of RGT Assessment}
\label{rq2-sec}

Per the protocol described in \autoref{subsec:protocols}, we identify false positives of RGT assessment by manual analysis of the patches where at least one RGT test fails. Over the 257 patches from $D_{correct}$, 
RGT patch assessment yields 6 false positives. 
\emph{This means the false positive rate of RGT assessment is $6/257=2.3\%$} (\#F3).

We now discuss the 6 cases that are falsely classified as overfitting by RGT assessment.
They are classified into four categories according to the root causes and described in the first column  in  \autoref{tab:rq2-false-positive}. The second column presents the patch name, the third column  shows the category of behavioral difference as defined in \autoref{tab:rgt-oracles}. The fourth column gives the RGT test set that contains the failing test and the last column gives a short explanation.

 \begin{table*}[!t]
  \centering
  \notsotiny
  \caption{False Positive Cases by RGT Assessment}	
  \label{tab:rq2-false-positive}
  \renewcommand{\arraystretch}{1.3}
   \begin{tabularx}{\textwidth}{p{0.12\linewidth}p{0.26\linewidth}p{0.06\linewidth}p{0.08\linewidth}p{0.33\linewidth}}
    \toprule
   Category & Correct Patches &Category & RGT & Reasons in Detail \\
    \midrule
  
  \multirow{2}{*}{PRECOND} &\multirow{2}{*}{patch1-Math-73-Arja}& $D_{exc2}$& \multirow{2}{*}{$Evos_{2019}$} & RGT samples  inputs  violate  implicit preconditions of the program\\
    \hline
    
    \multirow{2}{*}{EXCEPTION} & patch1-Lang-7-DeepRepair&  \multirow{2}{*}{$D_{exc\_loc}$} & \multirow{2}{*}{$Evos_{2019}$} & \multirow{1}{*}{Same exception thrown from }\\
 & patch1-Lang-7-ACS& && different functions\\
   \hline
   
 \multirow{2}{*}{ OPTIM} &\multirow{2}{*}{patch1-Math-93-ACS }& \multirow{2}{*}{$D_{assert}$} & \multirow{2}{*}{$Rand_{2019}$} &  The ground-truth patch is more precise than the APR patch.\\
 
  \hline
 
 \multirow{2}{*}{IMPERFECT} & patch1-Chart-5-Arja& \multirow{2}{*}{$D_{exc2}$}& \multirow{2}{*}{$Evos_{2019}$} & RGT reveals a limitation \\
 & patch1-Math-86-Arja& & &in  the ground-truth patch \\

 \bottomrule
      \end{tabularx}	
\end{table*}

\textbf{PRECOND}
The patch from \textit{patch1-Math-73-Arja} is falsely identified as overfitting because RGT samples inputs that violate implicit preconditions of the program (\#F4).
\autoref{rq1-casestudy-math73-arja} gives the ground truth patch, the Arja patch and the RGT test that differentiates the behavior between the patches.
In \autoref{casestudy-math73-c}, we can see that RGT samples a negative number \textit{\small-1397.1657558041848} to update the variable \texttt{\small functionValueAccuracy}. However, the value of \texttt{\small functionValueAccuracy} is used to compare  absolute values (see the first three lines of  \autoref{casestudy-math73-a}). 
It is meaningless to compare the absolute values with a negative number, an implicit precondition is that  \texttt{\small functionValueAccuracy} must be positive, but there is no way for the test generator to infer this precondition.

This case study illustrates that RGT patch assessment may create false positives because the used test generation technique is not aware of preconditions or constraints on inputs. This confirms the challenge of Evosuite for sampling undesired inputs \citep{challenge-evosuite-icst13-Fraser}. 
On the contrary, human developers are able to guess the range of acceptable values based on  variable names and common knowledge.  This warns us that better support for preconditions handling in test generation would help to increase the reliability of RGT patch assessment.

\begin{listing}
\caption{The Case Study of Patch1-Math-73-Arja }
\tiny
\noindent\begin{minipage}[b]{0.9\textwidth}
    \begin{lstlisting}[firstnumber=106]
      if (Math.abs(yInitial) <= functionValueAccuracy){...}
      if(Math.abs(yMin) <= functionValueAccuracy){...}
      if (Math.abs(yMax) <= functionValueAccuracy){...}
     <@\colorbox{lightgray}{+   if (yMin * yMax > 0) \{ }@>  
     <@\colorbox{lightgray}{+      throw MathRuntimeException... \} }@> 
   
    \end{lstlisting}
    \subcaption{The ground truth patch for Math-73.}
    \label{casestudy-math73-a}           
    \end{minipage}%
    \hfill
    \noindent\begin{minipage}[b]{0.9\textwidth}
    \begin{lstlisting} [firstnumber=136]
       if (Math.abs(yMax)<=functionValueAccuracy{...}
     <@\colorbox{lightgray}{+ verifyBracketing(min, max, f);}@> 
       return solve(f,min,yMin,max,yMax,initial,yInitial);
    
    \end{lstlisting}
    \subcaption{The generated  patch of patch1-Math-73-Arja.}
    \label{casestudy-math73-b}            
    \end{minipage}%
    \hfill
\begin{minipage}[b]{0.9\textwidth}
    \begin{lstlisting}[firstnumber=665]
       double double1 = <@\colorbox{lightgray}{-1397.1657558041848 }@> ;
       brentSolver0.setFunctionValueAccuracy(double1);
    \end{lstlisting}
    \subcaption{The generated test that fails on the generated patch.}   
    \label{casestudy-math73-c}            
\end{minipage}

\label{rq1-casestudy-math73-arja} 
\end{listing}

\textbf{EXCEPTION}
Both \textit{patch1-Lang-7-SimFix} and \textit{patch1-Lang-7-ACS} throw the same exception as the one expected in the ground truth program: \textit{fail ("Expecting exception: NumberFormatException")}.  

However, these two patches are still assessed as overfitting because the exceptions are thrown from different functions from the ground truth program. Per the introduction of behavioral difference $D_{exc\_loc}$ in  \autoref{tab:rgt-oracles}, exceptions thrown by different functions justify an overfitting assessment.  

RGT assessment yields two false positives when verifying exceptions thrown positions.
This suggests that category $D_{exc\_loc}$ may be skipped for RGT, which is easy to adjust by configuring corresponding options in test generators.

\textbf{OPTIM} The \textit{patch1-Math-93-ACS} is assessed as an overfitting patch by $RGT_{Randoop2019}$ tests  because they detect behavioral differences of $D_{assert}$.
Bug Math-93 deals with computing a value based on logarithms. The  fix from ACS uses  $ln^{n!}$,  which is  mathematically equivalent to the human-written solution $\sum ln^{n}$. Their behavior should be semantically equivalent. 
However, the human-written patch introduces optimization for calculating  $\sum ln^{n}$ when $n$ is less than $20$ by returning a precalculated value.
For instance, one of the sampled input is \textit{n=10}, the expected value from the ground truth patch is \textit{\small 15.104412573075516d} (looked up in a list of hard-coded results), however, the actual value of \textit{patch1-Math-93-ACS} is \textit{\small 15.104412573075518d}. Thus, an assertion failure is caused and RGT classifies this patch as an overfitting patch because of such behavioral difference in output value.  This false positive case would have been avoided if no optimization was introduced in the human-written patch that was taken as a ground truth.  

Our finding warns the reproducible bug benchmark work (e.g., \citep{Bears2019, DefextsICSE19}) should 
 pay additional attention to distinguishing the optimization code from the repair code in the human-written (reference) patches (\#F5).

  \begin{listing}
  \caption{A Null Pointer Exception Thrown in Assessing Patch1-Chart-5-Arja}
  \label{chart-5}
   \begin{lstlisting}[firstnumber=593] 
     for (int i = 0; i < this.data.size(); i++) {
        XYDataItem item = (XYDataItem) this.data.get(i);
        if (<@\colorbox{lightgray}{item.getX().equals(x)}@>) {

    \end{lstlisting}
 \end{listing}

\textbf{IMPERFECT}
Two cases  are falsely classified as overfitting due to the imperfection of human-written patches.  They both cause the behavioral difference category $D_{exc2}$ that no exception is expected from a ground truth program while exceptions are thrown from a patched program.
The \textit{patch1-Chart-5-Arja} throws a null pointer exception because the variable \texttt{item} is null when executing RGT tests. The code snippet is given at line 595 of \autoref{chart-5}. The human-written patch returns earlier, before executing the problematic code snippet, while the fix by \textit{patch1-Chart-5-Arja} is later in the execution flow. 
Hence, an exception is thrown from \textit{patch1-Chart-5-Arja} but not from the human-written patch for the illegal input.
Another patch of \textit{patch1-Math-86-Arja} can actually be considered better than the human-written patch because it is able to signal the illegal value \texttt{NAN} by throwing an exception while the human-written patch silently ignores the error (\#F5).

\emph{Is the human written patch a perfect ground truth?}
RGT and related techniques are based on the assumption that the human-written patches are fully correct. Thus, when a test case differentiates the behavior between an APR patch and a human-written patch, the APR patch is considered as overfitting.
The experimental results we have presented confirm that human-written patches are not perfect.
Our findings confirm that the human patch itself may be problematic \citep{Gu2010BRF, Yin-fse11}. However, we are the first to reveal how the imperfection of human patches impacts automatic patch correctness assessment.
Beyond that, as shown in this section, optimization introduced at the same commit of bug fixing and other limitations influence overfitting patch identification of RGT assessment.

\begin{mdframed}
Answer to RQ2: 
According to this experiment, the false positive rate of RGT patch assessment is 6/257 = 2.3\%. Considering this false positive rate as reasonable, the program repair researchers can rely on this technique for providing assessing results of their program repair contributions, this  automated assessment being more reliable than manual patch assessment. Moreover, our detailed case studies show that blindly considering the human-written patch as perfect ground truth is wrong, some corner-cases exist. To our knowledge, this is the first analysis of the false positives for automated patch assessment.
\end{mdframed}

\begin{figure}[!t]
\caption{The Effectiveness of RGT  and DiffTGen } 
\includegraphics[width=0.96\linewidth]{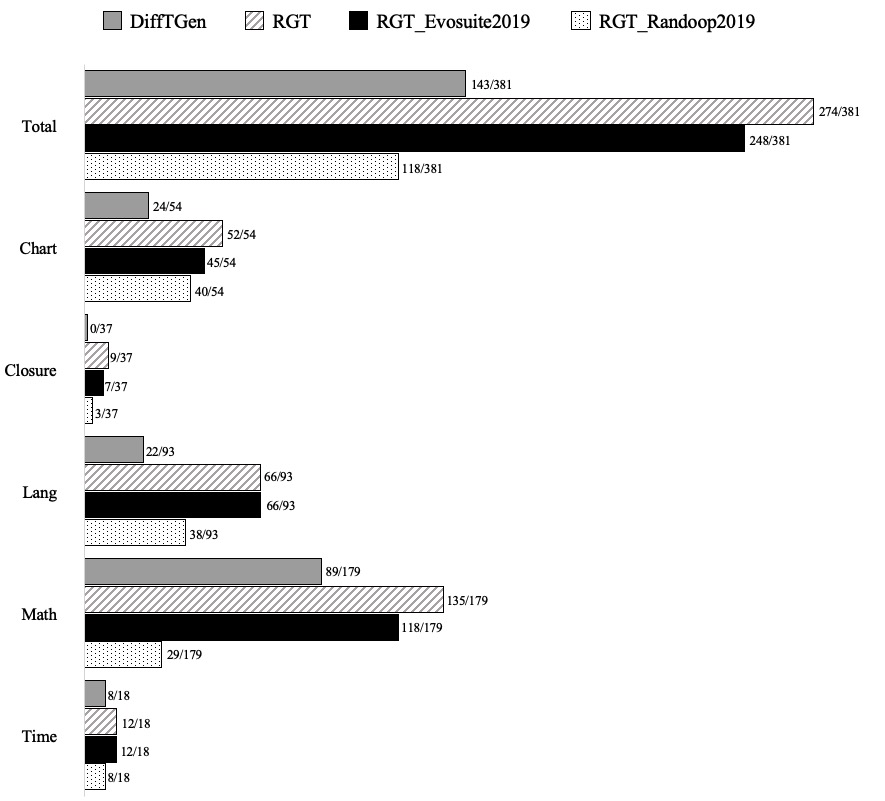}
\label{fig-rq3}
\end{figure}

\subsection{Result of RQ3: Effectiveness of RGT Assessment Compared to DiffTGen}
\label{rq3-sec}

\subsubsection{The Effectiveness of RGT Assessment}
We have executed 30 runs of DiffTGen over  $D_{correct}$.
DiffTGen identifies 2 patches as overfitting, which were both misclassified as correct (\textit{patch2-Lang-51-Jaid} and  \textit {patch1-Math-73-JGenProg2015}).
Recall that RGT patch assessment identifies in total 10 misclassified patches, including the 2 mentioned patches found by DiffTGen. 
This shows that RGT is more effective than DiffTGen.

Per the core algorithm of DiffTGen and its implementation, DiffTGen can only handle category $D_{assert}$ of behavioral difference (value difference in the assertion).
However, DiffTGen fails to identify another two misclassified patches also found by RGT of $D_{assert}$ category: \textit{patch1-Lang-58-Nopol2015} and  \textit   {patch1-Lang-41-SimFix}. 
Because DiffTGen fails to sample an input that differentiates the instrumented buggy and human-written patched programs, while our RGT assessment does not require those instrumented programs.

Further, we have performed 30 executions of RGT tests and DiffTGen over the whole 381 patches from $D_{overfitting}$. 
$RGT_{Evosuite2019}$ yields \numprint{7923} test failures and $RGT_{Randoop2019}$ yields \numprint{65819} test failures.  
Specifically, $RGT_{Evosuite2019}$ identifies  248 overfitting patches and $RGT_{Randoop2019}$  identifies 118 overfitting patches, and together they identify 274 overfitting patches (\#F6).
DiffTGen identifies 143/381  overfitting patches.  
Our experiment provides two implications: (1) RGT patch assessment improves over DiffTGen, and (2) For RGT patch assessment, Evosuite outperforms Randoop in sampling inputs that differentiate program behaviors by 210\% (248/118), but consider these two techniques together can maximize the effectiveness of overfitting patches identification (\#F7).

\autoref{fig-rq3} shows the number of overfitting patches in $D_{overfitting}$ dataset identified by RGT assessment and DiffTGen. RGT gives better results than DiffTGen for all Defects4J projects.  
An outlier case is the Closure project, for which we see that the assessment effectiveness is low, both for RGT (9/37) and for DiffTGen (0/37). 
This is consistent with the results as shown in \autoref{fig:coverage}: RGT tests generated for the Closure project have the lowest coverage.
As a result, the sampled RGT tests are less effective in exposing behavioral differences in the Closure project.

\begin{figure}[!t]
\caption{RGT Code Coverage for Generated Test Suites that Detected Overfitting Patches and Not Detected.} 
\includegraphics[width=0.97\linewidth]{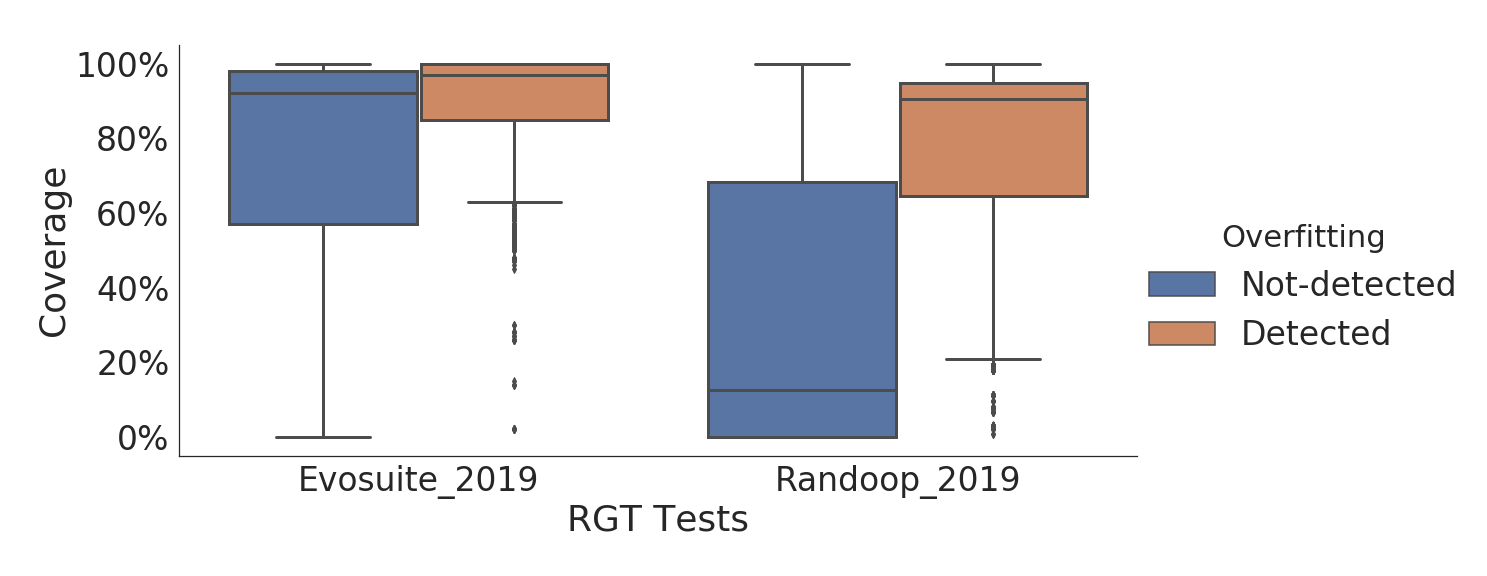}
\label{coverage-relation}
\end{figure}

\subsubsection{On Patch Assessment and Code Coverage}

\autoref{coverage-relation} compares the code coverage obtained by the RGT test suites that detect overfitting patches and against the ones that do not detect overfitting patches. 
It shows that the test suites that detect overfitting patches have higher code coverage. 
Indeed, the average code coverage is 84\% for tests that detect overfitting patches and 51\% for the rest.
In addition, we conduct a Mann-Whitney U test  \citep{Mann-whitney-utest} to confirm that the difference between these two categories is significant, which is the case, the p-value is lower than $0.001$.
This shows that the RGT tests with higher test coverage are more likely to expose program behavioral differences and to detect overfitting patches for program repair.


\subsubsection{On the Difference between RGT and DiffTGen}

\begin{figure}[!th]
\caption{The Proportion of Behavioral Difference Categories } 
\includegraphics[width=0.96\linewidth]{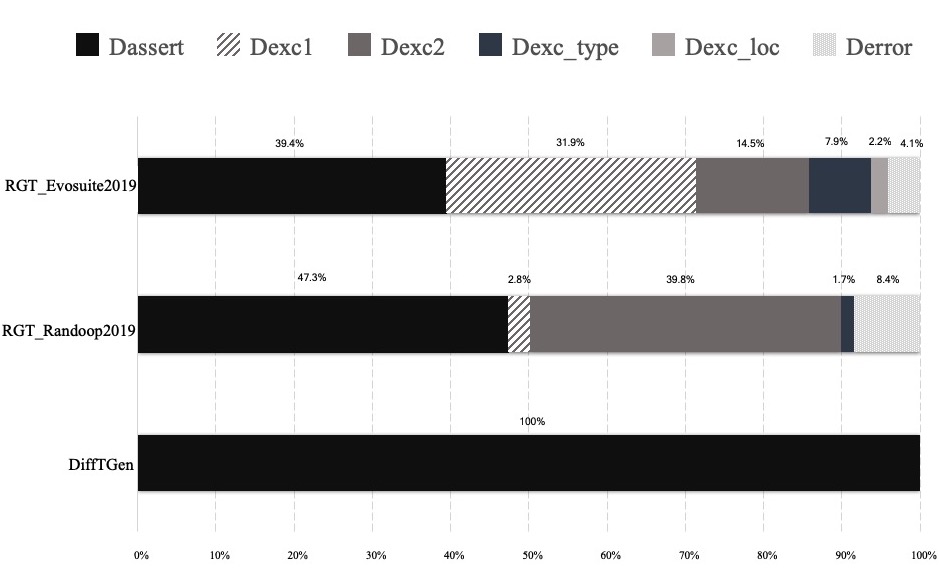}
\label{fig-rq3-diff}
\end{figure}

\autoref{fig-rq3-diff} shows the proportion of behavioral differences detected by RGT tests and DiffTGen per the taxonomy presented in \autoref{tab:rgt-oracles}.
The proportions are computed over
\numprint{7923} test failures of $RGT_{Evosuite2019}$,  \numprint{65819} test failures of $RGT_{Randoop2019}$, and 143 behavioral differences detected by DiffTGen. 
$RGT_{Evosuite2019}$ (top horizontal bar) detects six categories of behavioral differences and 
$RGT_{Randoop2019}$  detects five categories.  DiffTGen is only able to detect behavioral differences due to assertion failure between expected and actual values. 
For example, DiffTGen fails to produce a result for the two Lang-43 patches shown in \autoref{case-study-lang-43}. The reason is that these two patches cause a Java heap space error, thus no values are produced for comparison in DiffTGen. On the contrary, RGT works on these cases, it can successfully compare the behavioral difference and detect these two overfitting patches.

In all cases, we see that assertion failure is the most effective category to detect behavioral differences of overfitting patches.  
Moreover, exceptions are also effective to detect behavioral differences, and this is the key factor for RGT's effectiveness over DiffTGen (\#F8).
Notably, the two considered test generators are not equally good at generating exceptional cases, e.g., 31.9\% of $RGT_{Evosuite2019}$ failing tests expose differences of category $D_{exc1}$ while only 2.8\% of $RGT_{Randoop2019}$ tests do so. Similarly, we note that Randoop does not support exception assertions based on the thrown location ($0\%$ of $D_{exc\_loc}$).

\EMSErevision{ Recall that both DiffTGen and $RGT_{Evosuite2019}$ leverage Evosuite for test case generation, now we explain how we obtain those differences based on the configuration difference. We present the Evosuite configurations in \autoref{confi-rgt-difftgen}. 
The first column gives the parameter to configure Evosuite, the second and third columns show the value set for such a parameter by DiffTGen and RGT respectively, and we explain the meaning of such a parameter in the last column for clarification. 
All other parameters are set to their default value. 
}

\EMSErevision{
As shown in \autoref{confi-rgt-difftgen}, both DiffTGen and $RGT_{Evosuite2019}$ configure the search criterion to \textit{branch} coverage to guide Evosuite to generate tests, i.e., it maximizes branch coverage. The second row indicates that they both execute Evosuite 30 trials with 30 different random seeds.
DiffTGen considers 60 seconds for the search budget (the best configuration of DiffTGen reported in \cite{issta17-difftgen}).
$RGT_{Evosuite2019}$ considers 100 seconds for the searching budget which is heuristically the best value for RGT we identified in our experiments. 
DiffTGen does not configure a timeout for executing the body of a single test. On the contrary, RGT configures a timeout to bound the experimental time.  
}
\EMSErevisionMinor{As shown in \autoref{fig-rq3-diff},  no overfitting patch is identified by $RGT_{Evosuite2019}$ with the timeout difference (e.g., $D_{timeout}$). In other words, the timeout difference setting has no influence on the experiment results. 
Thus, the experimental evaluation can be considered fair.
}

\EMSErevision{
DiffTGen and $RGT_{Evosuite2019}$ differ in one key parameter: assertion generation.
DiffTGen configures the assertion as \textit{false} in Evosuite because it does not compare the behavior based on the oracles generated by Evosuite but based on the variables observed via monitoring with code instrumentation. Recall that to determine a patch's correctness, DiffTGen compares the values of instrumented variables between the patched version and the human-written version. 
On the contrary, RGT fully leverages the oracles (i.e., assertions and exceptions) generated by Evosuite based on the human-written version. 
In summary, DiffTGen and $RGT_{Evosuite2019}$ use the same search criterion, random seeds, and close search budget to guide Evosuite for test generation.  
This key difference between DiffTGen and $RGT_{Evosuite2019}$ is the generation of assertion: $RGT_{Evosuite2019}$ uses Evosuite to generate executable test cases with oracles while DiffTGen only considers differences in internal variables.
}

 \begin{table}[t!]
 \centering
 \footnotesize
 \renewcommand{\arraystretch}{1.4}
\caption{\EMSErevision{Configerations of Evosuite in DiffTGen and $RGT_{Evosuite2019}$}}
 \label{confi-rgt-difftgen}
 \begin{tabular}{lccl}
 \hline
Parameters & DiffTGen& $RGT_{Evosuite2019}$ & Description of parameters\\
\hline
Criterion & branch & branch &  Coverage criterion\\
 \hline
Random seeds  & 30 & 30 &  The times of Evosuite are invoked \\ 
\hline
Search budget& 60 & 100   & Search duration for test generation  \\ 
\hline
Timeout  & - & 3000 & Milliseconds allowed to execute a test \\
\hline
Assertion&false & true  & Create assertions \\

\hline
\end{tabular}
\end{table}


We also compared the ability of DiffTGen and $RGT_{Evsouite2019}$ to capture the output differences. 
DiffTGen captures the results of the execution of each statement (if any) and then compares, for each statement, the result obtained from the human-written patch and that from the machine patch(per the design of DiffTGen, these oracles are usually manually constructed)).
Due to its design, DiffTGen requires the compare values that are present in both the human-written patch and the machine patch.
In our experiments, DiffTGen fails to capture all output differences for two reasons:
1) there are no instrumented output values available, or 
2) the output values are not comparable. 
For example, DiffTGen fails to capture 16 overfitting patches generated for bug Chart-1, because neither the faulty program line nor the patched program line is a value line, and thus no output values are captured. 
On the contrary, $RGT_{Evsouite2019}$ tests consider all possible variables in generated assertions. $RGT_{Evsouite2019}$ captures more behavioral differences by exploring all possible variables as well as more properties of those variables.

\EMSErevision{
We recapitulate the main novelty and advantages of RGT compared to DiffTGen.
First,  RGT provides reusable tests that can be executed in a lightweight manner on any machine patches. On the contrary, in DiffTGen, all tests are generated based on an instrumented patched program, and these tests are coupled with the specific instrumented variables. Thus, the generated tests of DiffTGen are not reusable for future research. 
Second,  RGT is a fully automated technique while DiffTGen requires manual work to identify change-related statements in the patched version and the human-written version (this has also been noted by \citet{XiongIdentifyingPC-ICSE18-patchsim}). 
}

\EMSErevision{ Now we compare our findings against those of the close related work by  \citet{le:reliability-patch-assess}.
First of all, both experiments find that the performance of DiffTGen and Randoop for detecting overfitting patches is similar. Since, our experiment is done on a new and bigger benchmark (381 versus 135 patches overfitting patches), this significantly increases the external validity of this finding.\footnote{As aforementioned, our study and \citet{le:reliability-patch-assess} both configure DiffTGen with its best configuration of 30 trails and 60 seconds for the search budget. 
Our study configures Randoop to run 100 seconds while they configure Randoop to run 3 minutes, yet this different configuration of the search budget does not bring much difference (per \citep{ase15-tests} and our own experience).  Randoop's code coverage is saturated already within 60 seconds and after that, the generated test suites exhibit a very high degree of redundancy. }
Second, the key novelty of our experiment is that we consider Evosuite which is not used in \citet{le:reliability-patch-assess}.
In our experiment, DiffTGen and Randoop respectively achieve the effectiveness of 37.5\%, 31\%, while Evosuite reaches 65.1\%. 
This is a major result compared to \citet{le:reliability-patch-assess}: it shows that automated patch assessment is actually effective, which is essential for future progress in program repair.
Finally, we suggest that different test generation tools can be used in combination, which is a pragmatic approach for practitioners: our study shows that Evosuite and Randoop put together in RGT can achieve a 72\%  effectiveness in identifying overfitting patches.
}

\begin{mdframed}
Answer to RQ3: 
Out of the 381 patches claimed as overfitting by manual analysis, RGT assessment automatically identifies 274 / 381 of them (72\%). 
This is a significant improvement over \citet{le:reliability-patch-assess} which reported that fewer than 20\% of overfitting patches could be identified.
RGT improves the state-of-the-art technique DiffTGen by 190\% (274 versus 143 patches detected as overfitting), which strongly signals that RGT can reliably alleviate researchers to manually label overfitting patches. 
Our experiment is notable by its scale: the most related experiments are \citet{le:reliability-patch-assess} and \citet{issta17-difftgen}  based on 135 and 79 overfitting patches respectively, as opposed to 381 in this study (thus our experiment is performed on a 2.8X and 4.8X larger dataset). 

\end{mdframed}


\begin{table*}[!t]
  \centering
  \scriptsize
   \renewcommand{\arraystretch}{1.4}
  \caption{Time Cost of RGT Patch Assessment}
   \label{rq4-time-cost}
	\begin{tabular}{lccccccc}
    \toprule
    & \multicolumn{5}{c}{RGT} \\
     \cline{2-6}
    &\tiny{$Evos_{2019}$}&\tiny{$Rand_{2019}$}& \tiny{$Evos_{ASE15}$ }&   \tiny{$Rand_{ASE15}$} &  \tiny{$Evos_{EMSE18}$}  \\
    \hline
    TCGen&136.3 hrs &109.7 hrs   &-&-&-\\
    \hline
   SC& 2.9 hrs&2.5 hrs&1.3 hrs &2.6hrs &1.1 hrs \\
    \hline
    EXEC1 on $D_{correct} $& 6.2 hrs&5.2 hrs & 1.6 hrs & 5.1hrs &1.7hrs \\
    EXEC2 on $D_{overfitting}$ & 9.1 hrs & 7.7 hrs& 2.3 hrs & 7.6hrs & 2.3hrs \\
    \hline
    sum in hours &154.5 hrs&125.1 hrs  &5.2 hrs & 15.3 hrs &5.1 hrs \\
    
    \bottomrule
    \end{tabular}
    \end{table*}
    

\subsection{Result of RQ4: Time Cost of RGT Patch Assessment}
\label{rq4-sec}

\autoref{rq4-time-cost} summarizes the time cost of RGT patch assessment.
The first column gives the breakdown of time cost as explained in \autoref{subsec:protocols}.  The second and third columns give the cost for the RGT tests we have generated for this study, while the fourth to sixth columns are the three categories of RGT tests generated in previous research projects shared by their respective authors.
$TCGen$ time is not available for the previously generated RGT tests.  They were reported by their authors but it is not our goal, thus we put a `-` in the corresponding cells.
For example, the second column indicates $RGT_{Evosuite2019}$ required 
136.3 hours for test case generation, 2.9 hours for performing the sanity check, and 6.2 hours for assessing the correctness of patches in $D_{correct}$ dataset and 9.1 hours in $D_{overfitting}$ dataset.

We observe that $TCGen$ is the dominant time cost of RGT patch assessment. $RGT_{Evosuite2019}$ and $RGT_{Randoop2019}$ respectively spend 
136.3/154.5 hours (88.2\%) and 109.7/125.1 hours (87.7\%) on test generation (\#F10). 
For assessing 638 patches using newly generated RGT tests, we need 14.5 minutes and 11.76 minutes per patch for Evosuite and Randoop respectively.

The three sets of previously generated RGT tests require $5.2$, $15.3$ and $5.1$ hours  in accessing patch correctness for $D_{correct}$ and $D_{overfitting}$ dataset. Our experiment presents reusing tests from previous research is a significant time saver. For assessing 638 patches using previously generated RGT tests, the assessment time is 2.4 minutes per patch on average.

Note that the execution time of $RGT_{EvosuiteASE15}$ is less than $RGT_{Evosuite2019}$.
 This is because $RGT_{EvosuiteASE15}$ contains only 10 runs of test generation but $RGT_{Evosuite2019}$ contains 30 runs. 
With the same number of test generation configurations, $RGT_{EvosuiteEMSE18}$ goes faster than $RGT_{Evosuite2019}$, because it only contains tests for 42 bugs.

Now we take a look at the effectiveness of RGT tests from previous research.
RGT tests generated from previous research identifies 9 out of 10 misclassified patches from $D_{correct}$ (the missing one is  \textit{patch1-lang-35-ACS}).
From $D_{overfitting}$, a total of 219 overfitting patches are found by the three previously generated RGT tests together  (\#F11).
Recall that $RGT_{Evosuite2019}$ and  $RGT_{Randoop2019}$ together identify 274 overfitting patches for $D_{overfitting}$. 
Despite a fewer number of tests, RGT tests  from previous research achieve 80\% (219/274) effectiveness compared to our newly generated RGT tests.  
Therefore, RGT tests generated from previous research are considered effective and efficient for patch correctness assessment usage.

Regression test generation is known to be costly and indeed, over 87\% of the time cost in our experiment is spent in test generation. Consequently, reusing previously generated RGT test cases is a significant time-saver for patch assessment. 
By sharing a curated dataset of 4 million generated RGT tests, we save 246 computation hours for future researchers (not counting the associated time such as configuration, cluster management, etc). More importantly, reusing tests is essential for the scientific community: when experiments and papers are based on the same set of generated tests, the results can be reliably compared one against the others. Consequently, our replication dataset helps the community to achieve well-founded results.

\begin{mdframed}
Answer to RQ4: 
Over 87\% of the time cost of RGT patch assessment is spent in test case generation. Consequently, it is recommended to durably share previously generated RGT tests for time-saving. This also greatly improves scientific reproducibility and coherence because all researchers can assess the APR patches on a given benchmark with the same generated tests.
\end{mdframed}

\begin{figure}[h]
\caption{The number of overfitting patches found depending on the number of test generations. The X-axis indicates the number of test suite generation and the Y-axis indicates the number of overfitting patches found. }
\includegraphics[width=0.97\linewidth]{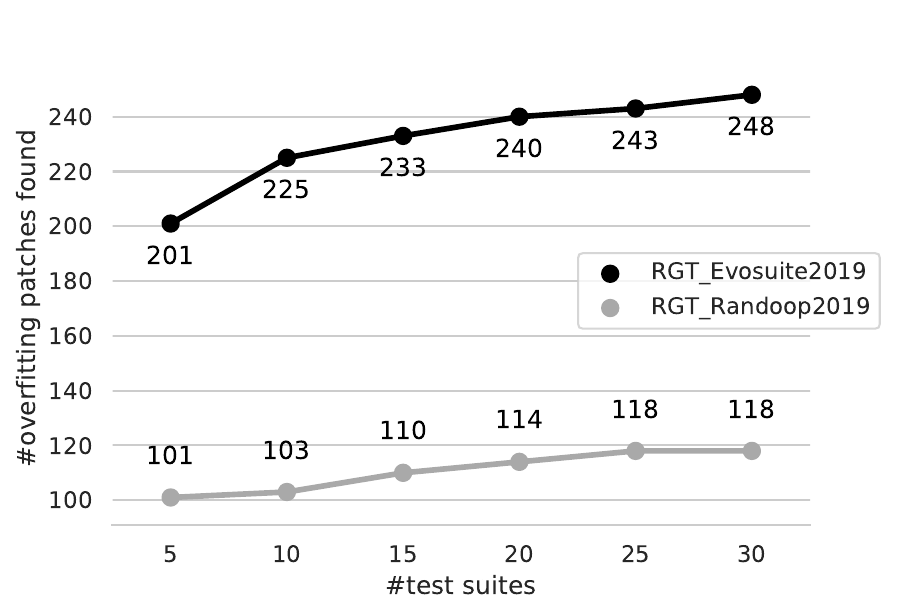}
\label{fig-rq5-tradeoff}
\end{figure} 
\subsection{Result of RQ5:  Trade-off between Test Generation and Effectiveness of RGT Assessment}
\label{rq5-sec}
 
\autoref{fig-rq5-tradeoff} shows the number of overfitting patches detected depending on the number of generated test suites. The X-axis shows the number of test suites generated with a different seed, the Y-axis is 
the average number of detected overfitting patches over 1000 random groups sampled from $D_{overfitting}$.
Recall that the best results, obtained after all runs, are that $RGT_{Evosuite2019}$ and $RGT_{Randoop2019}$  identify 248 and 148 overfitting patches individually from $D_{overfitting}$. 

For both $RGT_{Evosuite}$ and $RGT_{Randoop2019}$, the more test generation runs, the better the effectiveness of RGT patch assessment. 
Nevertheless, even with a small number of test generations, e.g., 5 runs, RGT is able to achieve more than 80\% of effectiveness.
On average, 25 runs of $RGT_{Randoop2019}$ is able to achieve the same performance as 30 runs of Randoop.  On the contrary, $RGT_{Evosuite}$ keeps identifying more overfitting patches, even after 25 runs. Due to the computational costs of this experiment, it is left to future work to identify when a plateau appears for $RGT_{Evosuite}$ . 
We observe that after 10 test suites of $RGT{Evosuite 2019}$, the newly identified overfitting patches still increase but do not largely vary. Thus, a pragmatic rule of thumb is to do 10 test generations. 
For $RGT_{Randoop2019}$, the number of overfitting patches identified by different numbers of test generation is considerably close.  In our experiment, we observe after 15 test suites in $RGT_{Randoop2019}$, the curve starts to remain steady. Thus, a pragmatic rule of thumb is to do 15 test generations in Randoop that is equivalent to  93\% effectiveness of overfitting patch classification.

\begin{mdframed}
Answer to RQ5: 
The more test suites generated, the better effectiveness of overfitting patch identification. Yet, our experiments suggest pragmatic values  to be used by APR researchers: 10 runs of Evosuite and 15 runs of Randoop (\#F12).
\end{mdframed}

\section{Actionable Data}

Table \ref{findings} at the beginning of this paper lists the actionable implications obtained with our original experiments.
Furthermore, our work provides actionable data for future research in automatic program repair. 

\emph{A dataset of 638 APR patches for Defects4J.} We have collected and canonicalized 638 original patches from 14 different repair systems that form our experiment dataset. All patches have gone through strict sanity checks. This is a reusable asset for future research in program repair in particular to study anti-overfitting techniques and behavioral analysis.

\emph{A dataset of \rgttotalnumber RGT tests for Defects4J.} We have curated \rgttotalnumber{} generated test cases from two test generation systems. They complement the manual tests written by developers with the new assertions and input points sampled from the input space.
Overall, they provide a specification for Defects4J bugs. Given the magnitude, it is possibly the largest specification ever of the expected behavior of Defects4J bugs. 
This is essential for program repair research which heavily relies on Defects4J. We believe it could be of great value as well in other research fields such as fault localization, testing and bug clustering.

\section{Threats to Validity}
\label{sec:threats}
We now discuss the threats to the validity of our results. 

\emph{Threats to internal validity}
A threat to internal validity relates to the implementation of the methodology techniques.
 1) Threats to  validity in RGT. The removal of flaky tests from RGT may discard test inputs that could expose behavioral differences. For this reason, the results we report are potentially an under-estimation of RGT's effectiveness.
 2) Threats to validity in DiffTGen. DiffTGen requires a mandatory configuration about syntactic deltas, which are not provided by the authors of DiffTGen. Consequently, in our experiment, we improved DiffTGen to  automatically generate the delta information. We observe that minor differences in those deltas could produce different results: this poses a threat to the DiffTGen results reported in this paper. 
 We provide the delta information in our public open-science repository \citep{repo} so that future research can verify them and build on top of them.

\emph{Threats to external validity} 
  The threats to external validity correspond to the generalizability of our findings. In this paper, we perform our experiments on the Defects4J benchmark with 638 patches. We acknowledge that the results may differ if another bug benchmark is used \citep{RepairThemAll2019,Bears2019}. Future research on other benchmarks will further improve the external validity. To the best of our knowledge, our experiment on analyzing 638 patches from automatic repair research with \rgttotalnumber{} generated tests are the largest ever reported.

\section{Discussion}

\begin{listing}

\caption{The Case Study of Two False Negative Cases}
\label{case-study-math-53}
\noindent\begin{minipage}[b]{0.9\textwidth}
    \begin{lstlisting} [firstnumber=150] 
     public Complex add(Complex rhs) throws NullArgumentException {            
     MathUtils.checkNotNull(rhs);            
     <@\colorbox{lightgray}{+      if (isNaN || rhs.isNaN) \{   }  @>        
     <@\colorbox{lightgray}{+          return NaN;  }  @>       
     <@\colorbox{lightgray}{+        \}  }  @>            
    return createComplex ...           
     }
                                     
    \end{lstlisting}
    \subcaption{The human-written patch for Math-53}
    \label{patch-math-53-ground-truth}            
    \end{minipage}%
    \hfill
\begin{minipage}[b]{0.9\textwidth}
    \begin{lstlisting}[firstnumber=150] 
      public Complex add(Complex rhs)  throws NullArgumentException {            
     <@\colorbox{lightgray}{+      if (isNaN || rhs.isNaN) \{   }  @>        
     <@\colorbox{lightgray}{+          return NaN;  }  @>       
     <@\colorbox{lightgray}{+        \}  }  @>    
     MathUtils.checkNotNull(rhs);      
    return createComplex ...           
     }
  
    \end{lstlisting}
    \subcaption{The generated patch of patch2-Math-53-Jaid and patch2-Math-53-CapGen }
    \label{patch-math-53-1}            
\end{minipage}

\end{listing}

\subsection{Manual Verification of RGT Assessment}
\label{dis-manual-assess}

Among 97 semantically equivalent patches assessed as correct by previous manual analysis, RGT yields 16 program behavioral different patches and 81 behavioral equivalent patches.
After manual analysis, we find that RGT tests found 10 true-positive cases (patches misclassified in previous research, see \autoref{rq1-sec}) and 6 false-positive cases (see  \autoref{rq2-sec}).

Due to the potential incompleteness of RGT tests, we conduct a post-study by manually analyzing the remaining 81 patches identified as behaviorally equivalent by RGT tests. 
The manual analysis process is done following the recommendations from previous research \citep{zhongxing-EMSE18} and works as follows: (1) if a patch only partially fixes the bug, it is deemed as \textit{Overfitting-A} and (2) if a patch fixes the bug in general but introduces a new bug, it is deemed as \textit{Overfitting-B}. As a result, our manual analysis gives us two  \textit{Overfitting-B} cases which are false negatives. The remaining 79 patches are true negative cases.

We now discuss the two false-negative cases, where the corresponding code snippet is given in \autoref{case-study-math-53}.
Both the human-written patch and the two APR patches insert the same conditional check, but they are inserted in different positions. The human-written patch (\autoref{patch-math-53-ground-truth})  inserts the check before the call to \textit{MathUtils.checkNotNull} (line:151) but the two APR patches insert before this method call. 
There is a behavior difference when \textit{rhs} is null. 
A new null pointer exception would be thrown when \textit{rhs} is null in the APR patched version.
However, any RGT test containing a program input that produces variable \textit{rhs} receives a null.


In our study, the RGT assessment missed two subtle false-negative cases. With the previously reported 10 positive cases, this gives us the true positive rate (i.e., recall) of 83.3\% (10/12), which shows the RGT assessment is effective.
During our manual assessment process, we found it is hard to determine the correctness of a semantically equivalent patch due to lacking unified assessment criteria.
The manual assessment result may be different with different criteria. This reflects the bias problem of manual assessment. 

Our case study in \autoref{case-study-math-53},  together with the aforementioned case study in \autoref{case-study-lang-43}, they warn researchers the sensitive location of automatic program repair patches is of the high relevance of their correctness.
One of the future works of program repair should towards precisely synthesized repair patches.

\subsection{\EMSErevision{On the Comparison of PATCH-SIM}}
\EMSErevision{
Now, we present data analysis to compare the effectiveness of RGT with PATCH-SIM, the comparison is made on the same patches from the common repair systems. In their experiment and our experiment, there are a total of 20 overfitting patches in common: 5 from ACS, 6 from JGenProg2015, 6 from Nopol2015, and 3 from HDRepair. 
Our dataset does not overlap much with the PATCH-SIM's dataset. This is because we consider 14 repair systems while they consider 6 repair systems,  all patches collected in our work were manually assessed by respective authors from the tools (i.e., patches and labels are both from the original paper), while the correctness labels of some patches from PATCH-SIM were judged by \citet{XiongIdentifyingPC-ICSE18-patchsim} (e.g., patches from Nopol2017). 
}

\begin{figure}[t!]
\caption{\EMSErevision{The Intersection of Detected Overfitting Patches between RGT and PATCH-SIM in Four Repair Tools}} 
\includegraphics[width=0.95\linewidth]{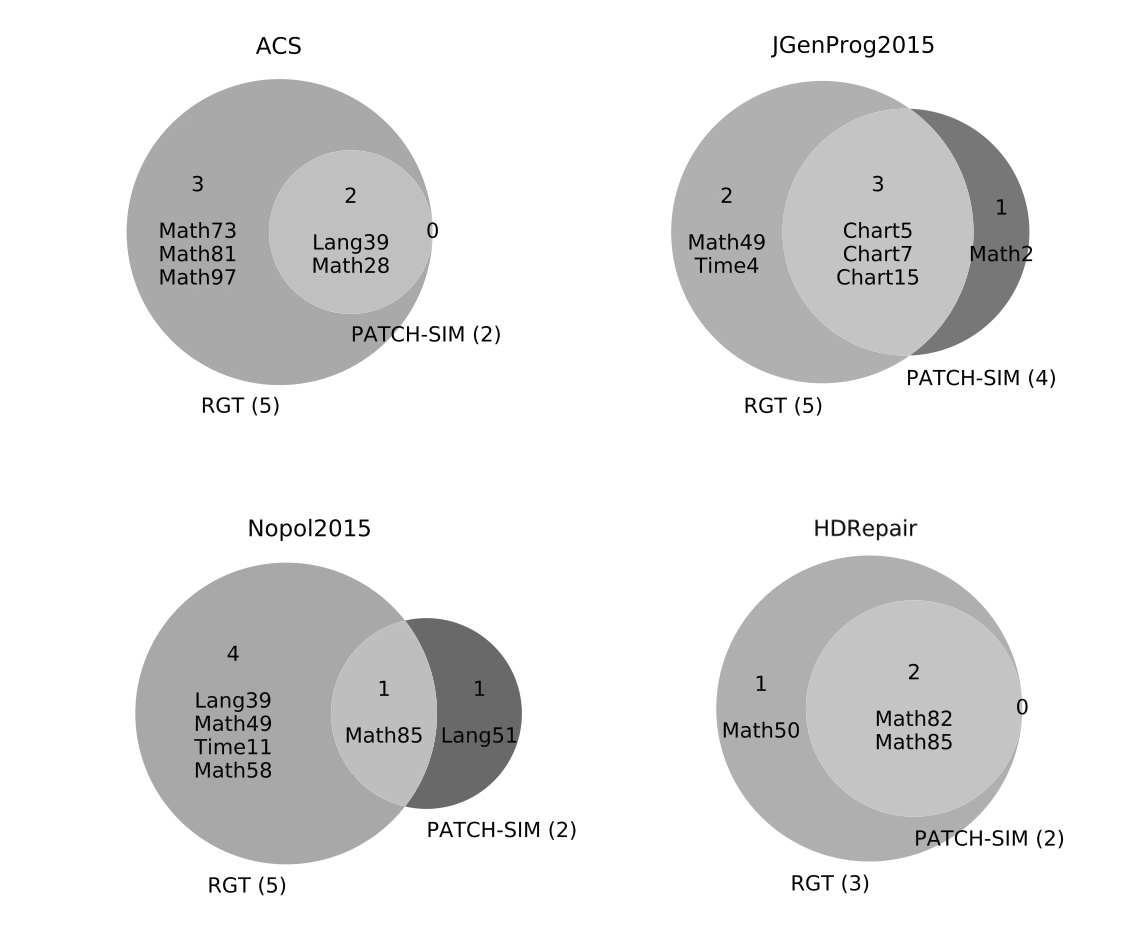}
\label{fig-PATCHSIM-RGT-Venn}
\end{figure}

\EMSErevision{
\autoref{fig-PATCHSIM-RGT-Venn} presents the intersection of detected overfitting patches between RGT and PATCH-SIM from these four repair tools as Venn diagrams.  The light gray indicates the overfitting patches identified only by RGT while the dark gray color indicates overfitting patches identified only by PATCH-SIM, the middle circles present the overfitting patches identified by both techniques.
For example, the Venn diagram in the top left corner shows there are 5 overfitting patches identified by RGT, with 2 of 5 can be identified by both RGT and PATCH-SIM, and no patch can be only detected by PATCH-SIM. 
In total, RGT identifies 18/20 overfitting patches while PATCH-SIM detects 10/20 patches. 
There is an overlap between the patches found by RGT and PATCH-SIM: 8 patches found by PATCH-SIM can also be identified by RGT.  In addition, there are 10 overfitting patches only found by RGT, and 2 overfitting patches only found by PATCH-SIM. Overall, this suggests that RGT  outperforms  PATCH-SIM in identifying overfitting patches. 
}

\subsection{On the Choice on Test Generators for RGT}
Many techniques have been proposed for automatic test generation, such as  Evosuite \citep{evosuite}, Randoop \citep{randoop}, T3 \citep{T3}, Sushi \citep{sushi}, TestFul \citep{testful}, eToc \citep{etoc} and Agitar\footnote{http://www.agitar.com, accessed on February 2020}.
According to the recent automatic test generation competition \citep{automatic-test-competition}, Evosuite and Randoop outperform T3 \citep{T3} and Sushi \citep{sushi}. 
We did not consider eToc, TestFul and Agitar because:
1) eToc is an old tool that has not been updated for several years,
2) TestFul and Agitar are not fully automated and require manual effort to generate a test suite.
In our study, we choose  Evosuite and Randoop because of their effectiveness and full support for automation.

In our experiment, we consider branch coverage to guide RGT test generation, because it has been reported as one of the most effective criteria \citep{test-ibf20,coverage-criteria}. We note that recent research \citep{test-ibf20} suggests that  weak mutation coverage and direct branch coverage can be used as the supplement of the branch coverage. It is future work to study how they can be used to increase the effectiveness of RGT tests to differentiate program behavior.

\subsection{On the Relation with Test Minimization, Selection and Prioritization}
RGT patch assessment executes all generated tests.
However, this could have some overhead due to the existence of redundant tests (i.e.,  identical or equivalent tests can be generated). Also, some tests do not cover the patched code. 
In the future, we plan to improve RGT assessment by considering techniques for test minimization, selection, and prioritization applied to generated tests \citep{regressionSurvey} to speed up patch assessment for program repair. We now discuss those techniques.


Test suite minimization techniques \citep{test-minimisation1, test-minimisation2} can enrich our work by eliminating redundant test cases.  For example, RGT assessment could benefit from \citep{hierarchy-minimisation} that derives a hierarchy of tests.


Test case selection techniques \citep{test-selection-1,test-selection-2,test-selection-3,test-selection-4} focus on identifying test cases that are relevant to a given change, in our case that is relevant to the patched code in the APR patches.
It is an interesting research direction to apply test case selection techniques to reduce RGT assessment costs by only selecting and running the tests that may be affected by the APR code changes.
We speculate that there is a trade-off between the effectiveness in differentiating program behaviors and the number of selected tests depending on the considered granularities \citep{hybrid-test-selection}.

Test case prioritization techniques \citep{test-prio-3,test-prio-2} rank tests by the likelihood of detecting problems.
Test case prioritization could support RGT patch assessment to
order test cases in such a way that overfitting patch detection is maximally effective. The top prioritized RGT tests can be executed first.
Overall, by augmenting RGT patch assessment with test suite minimization, selection and prioritization, it is likely that we can reduce the time cost of  RGT assessment.


\section{Related Work}
\label{sec:rw}
We now discuss the related work on patch correctness assessment and approaches focusing on alleviating overfitting patch generation.
\subsection{Patch Assessment}

To assess a patch, it is required to be able to cover the patch. Marinescu and Cadar~\citep{katch-fse13} proposed KATCH which uses symbolic execution to generate
test inputs that are able to cover the patched code. In our paper, we consider search-based test generation instead of symbolic execution approach.

The work most related to our paper is the study by  \citet{le:reliability-patch-assess}. In their study, they investigate the reliability of manual patch assessment and automatic patch assessment with DiffTGen and Randoop. 
There are four major differences between \cite{le:reliability-patch-assess} and our experiment:
1) our key result shows that 72\% of overfitting patches can be discarded with automated patch assessment, this is a significant improvement over \cite{le:reliability-patch-assess} in which fewer than 20\% of overfitting patches could be identified.
2) we provide novel experiments to comprehensively study automatic patch correctness assessment, including false positive measurement, time cost estimation, and trade-off analysis;
3) they consider patches generated by 8 repair systems while we consider 14 repair systems;
4) their dataset is composed of 189 patches while our dataset contains 638 patches.

\citet{quixbugs} use RGT tests to access patch correctness on QuixBugs benchmark.  There are two major differences in our experiment:
(1) their experiment is performed on small buggy programs in which the total amount of lines of code ranges from 9 to 67 lines. Our experiment is performed on real-world bug repositories.
(2) their dataset is composed of 64 patches while our dataset contains 638 patches.

There are several works focusing on alleviating overfitting patch generation from the perspective of practical usage, which is not an automatic patches correctness assessment for scientific study.

\citet{XiongIdentifyingPC-ICSE18-patchsim} propose PATCH-SIM and TEST-SIM to heuristically determine the correctness of
the generated patches without oracles. They run
the tests before and after patching the buggy program and measure the degree of behavior change. TEST-SIM complements PATCH-SIM by determining the test results of newly generated test inputs from Randoop. Our experiment shows Evosuite outperforms Randoop to sample test inputs to differentiate program behaviors. This suggests the effectiveness of this approach could be improved by also considering Evosuite for test generation.

Although PATCH-SIM is able to filter out overfitting patches, we consider RGT assessment is better than PATCH-SIM for scientific study for two reasons: 
(1) RGT assessment achieves better effectiveness of identifying overfitting patches (72\% of RGT and 56\% of PATCH-SIM reported in \citet{XiongIdentifyingPC-ICSE18-patchsim});
(2) PATCH-SIM suffers an 8.25\% false positive rate while RGT assessment reduces such a false positive rate to 2.3\%.
Recall that these two techniques have different goals. PATCH-SIM  targets APR users who do not have any ground truth patches, while RGT targets APR researchers who have ground truth patches, and help them assess better patch correctness with better scientific validity.
However, this technique could be improved by comparing the test execution difference with a ground truth program for scientific study. 
Nevertheless, due to the high cost of execution traces comparison of PATCH-SIM, this approach is too expensive for scientific patch assessment.



 \citet{anti-pattern} aim to identify the overfitting patches with the predefined templates to capture typical overfitting behaviors. They propose anti-pattern to assess whether a patch violates specific static structures. Recent work by \citep{ghanbari2019validation} aims to improve anti-pattern by combining it with machine learning techniques.
On the contrary,  RGT assessment fully relies on program run time behavioral differences to identify overfitting patches. 
While related, anti-pattern is not considered for assessing patch correctness.  Based on their static structures, 
the syntactically different yet semantically equivalent patches are typically not discarded with anti-patterns, as discussed by the authors.

\citet{Opad} propose Opad and  \citet{ISSTA19GaoCrash} propose Fix2Fit,  two approaches based on implicit oracles for detecting overfitting patches that introduce crashes or memory-safety problems.
Using these two approaches for automatic patch correctness assessment
would be an underestimation of overfitting patches, and also useless for Java where there is no memory problem.

\citet{qclose} propose Qlose to quantify the changes between the buggy program and the potential patch in terms of syntactic distances and semantic distances. 
They use program execution traces as a measure to rank patches.
With the ground truth patch, this technique can be used to assess the correctness of automatic repair patches.

In S3 \citep{s3}, the syntactic and semantic distances between a patched and a buggy program are used to drive synthesis for generating less overfitting patches. 
This approach could be extended with a ground truth patch to calculate the syntactic and semantic distances between an automatic repair patch and a ground truth patch for the usage of automated patch assessment.

Overall, all these techniques are overfitting patch identification techniques embedded in the repair process, they are not techniques for scientific evaluations of program repair research.

\subsection{Study of Overfitting}
\citet{cure-worse-15} find that overfitting patches fix certain
program behaviors, however, they tend to break otherwise correct behaviors. They study the impact of test suites coverage on generating correct patches:
test suites with higher coverage lead to higher quality patches. 
Consequently, patches generated with lower coverage test suites are prone to be overfitting.
Our study has a different scope, we look at the usage of generated tests for automatic correctness assessment, not the impact of coverage.

\citet{Long-search-space} conduct an analysis of the search spaces of two APR systems.  
Their analysis
shows that in the search space, there exist more overfitting patches than correct patches:
those overfitting patches that nevertheless pass all of the test cases are typically orders of magnitude more abundant. This presents the need for automated patch assessment technique. 
Our result of automatic patch correctness is encouraging news for researchers on accessing overfitting patches at scale. 

 \citet{QiIssta15-overfitting} and  \citet {Le:overfitting} perform empirical overfitting studies of automatic program repair.
They confirm automatic program repair indeed produces over 70\% to 98\% overfitting patches. 
By using RGT patch assessment, a  majority of manual work could be saved for APR patch correctness assessment.

\citet{zhongxing-EMSE18} analyze the overfitting problem in program repair and identify two overfitting issues: incomplete fixing (A-Overfitting) and regression introduction (B-Overfitting). The former one is about the fact that the generated patches partially repair the bug while the later one is about those patches which  break already correct behaviors.
Their experiments show that automatically generated tests are valuable to identify B-Overfitting(regression introduction). Our study to some extent confirms and complements their results. RGT tests based on regression oracles are effective to detect behavioral differences.
Their experiment is performed on 42 patches, our study has a much larger scope with an automated assessment of 638 patches (15 times bigger).

\subsection{Program Behavioral Difference Detection }

Comparing the behavior of programs is an important task in software development and maintenance. Different approaches have been proposed for this.


\emph{Delta debugging.}  \citet{DeltaAlgorithm} develop the delta comparison algorithm to systematically determine the minimal set of failure-inducing
changes between different program versions with given failure symptoms. 
Their approach requires the user to manually specify program locations where to compare internal program states. This technique could potentially be used in conjunction with RGT to minimize generated inputs.

\emph{Program spectra.}
Comparing execution spectra across program versions offer insights into the internal behavior differences \citep{SpectraRegression}.  
\citet{PathSpectra} compare path spectra from different runs of the program. 
PATCH-SIM is a typical program spectra approach to assess the correctness of a patch, which compares the execution trace differences between a buggy program and a patched program. This technique could potentially be used for patch assessment with ground truth execution traces provided by human-written patches.
\citet{prog-behav-diff-Xie} propose value spectra to expose program behavioral differences between versions even when their program outputs are the same.  
They collect internal program states at each user-function entry and exit as the value spectra of test execution. DiffTGen can be considered as a value spectra technique.


\emph{Invariants.} 
Program behaviors can be compared using invariants \citep{daikon}. 
\citet{invariant} propose an approach to execute a program on a collection of test cases against a collection of potential invariants. The inferred invariants capture program behaviors, and thus invariants can serve as test oracles to determine program correctness in newer versions.
It is an interesting direction of future work to compare the behavior of APR patches with invariants.

\emph{Symbolic execution.} Symbolic execution  explores all paths through a program to determine whether there are any program crashes or assertion violations. 
Differential symbolic execution (DSE) is proposed \citep{diff-symbolic-exe} to detect and characterize the effects of program changes in terms of behavioral differences between programs.
In theory, it is possible to use differential symbolic execution \citep{ICSE12-Symoblic-for-regression} to detect behavioral differences of APR patches.
However,  symbolic execution is known to have limitations to deal with program paths related to library code, structurally complex objects, strings, arrays, loops, etc. To our knowledge, symbolic execution in Java is not effective enough to be used for detecting behavioral differences in real-word complex programs such as the ones of Defects4J.


\emph{Implicit test oracles.} Implicit test oracles are those that rely on implicit knowledge to distinguish between correct program behavior and abnormal behavior. For instance, a crash or a buffer overflow should always be avoided. Fuzzing is an effective way to find violations of implicit oracles \citep{fuzz}. It has been shown that one can use implicit oracles for assessing patches \citep{ISSTA19GaoCrash,Opad}. 
However, patch correctness assessment with implicit oracles is mostly applicable to C code, and by nature provides an underestimation of the number of overfitting patches.

\section{Conclusion}
\label{sec:conc}
We have presented a large-scale empirical study of automated patch correctness assessment in this paper.
Our study confirms that manual patch correctness analysis is error-prone. Our automated patch assessment technique identifies 10 overfitting patches that were misclassified as correct by previous research.
All of them have been confirmed by the original authors (RQ1).
However, automated patch assessment is not completely perfect. We also measured a false positive rate of 2.3\% and discussed the false positive cases in detail (RQ2).
Overall, automated patch assessment is effective to identify 72\% overfitting patches, which saves much manual effort for APR researchers (RQ3).
Our experiment also shows that over 87\% time cost of RGT assessment is spent in test case generation (RQ4) and that
a trade-off exists between time cost in test generation and automated patch assessment effectiveness
(RQ5).

Our results are encouraging news for researchers in the program repair community: automatically generated test cases do help to assess patch correctness in scientific studies. To support the community and encourage automated patch assessment in future program repair experiments on Defects4J bugs, we make the dataset of 638 patches and \rgttotalnumber{} generated tests publicly available.

As future work, we will consider goal-directed test generation to cover the changes in patches. Also, it is promising to consider techniques on regression testing minimization, selection and prioritization \citep{regressionSurvey} in order to speed up patch correctness assessment. With such techniques, it would be possible to identify redundant test cases and to remove them from the RGT test suite, to prioritize test cases, in order to maximize the likelihood of detecting a behavioral difference in a given amount of time.

\section*{Acknowledgments}
We thank the anonymous reviewers for their insightful comments on the early draft of the paper.
This work was supported by the Wallenberg AI, Autonomous Systems and Software Program (WASP) funded by the Knut and Alice Wallenberg Foundation.

\bibliography{reference}

\begin{thebibliography}{76}
\providecommand{\natexlab}[1]{#1}
\providecommand{\url}[1]{\texttt{#1}}
\expandafter\ifx\csname urlstyle\endcsname\relax
  \providecommand{\doi}[1]{doi: #1}\else
  \providecommand{\doi}{doi: \begingroup \urlstyle{rm}\Url}\fi

\bibitem[Arcuri and Briand(2011)]{randomness-guide}
A.~Arcuri and L.~Briand.
\newblock A practical guide for using statistical tests to assess randomized
  algorithms in software engineering.
\newblock In \emph{Proceedings of the 33rd International Conference on Software
  Engineering}, ICSE '11, 2011.

\bibitem[Baresi and Miraz(2010)]{testful}
L.~Baresi and M.~Miraz.
\newblock Testful: Automatic unit-test generation for java classes.
\newblock volume~2, pages 281--284, 01 2010.

\bibitem[Barr et~al.(2015)Barr, Harman, McMinn, Shahbaz, and
  Yoo]{oracleproblem}
E.~T. Barr, M.~Harman, P.~McMinn, M.~Shahbaz, and S.~Yoo.
\newblock The oracle problem in software testing: A survey.
\newblock \emph{IEEE Transactions on Software Engineering}, 2015.

\bibitem[Benton et~al.(2019)Benton, Ghanbari, and Zhang]{DefextsICSE19}
S.~Benton, A.~Ghanbari, and L.~Zhang.
\newblock Defexts: A curated dataset of reproducible real-world bugs for modern
  jvm languages.
\newblock In \emph{Proceedings of the 41st International Conference on Software
  Engineering: Companion Proceedings}, ICSE '19, 2019.

\bibitem[Binkley(1995)]{test-selection-1}
D.~Binkley.
\newblock Reducing the cost of regression testing by semantics guided test case
  selection.
\newblock In \emph{Proceedings of the International Conference on Software
  Maintenance}, ICSM ’95, page 251. IEEE Computer Society, 1995.
\newblock ISBN 0818671416.

\bibitem[{Braione} et~al.(2018){Braione}, {Denaro}, {Mattavelli}, and
  {Pezzè}]{sushi}
P.~{Braione}, G.~{Denaro}, A.~{Mattavelli}, and M.~{Pezzè}.
\newblock Sushi: A test generator for programs with complex structured inputs.
\newblock In \emph{2018 IEEE/ACM 40th International Conference on Software
  Engineering: Companion (ICSE-Companion)}, May 2018.

\bibitem[Chen et~al.(2017)Chen, Pei, and Furia]{jaid}
L.~Chen, Y.~Pei, and C.~A. Furia.
\newblock Contract-based program repair without the contracts.
\newblock In \emph{2017 32nd IEEE/ACM International Conference on Automated
  Software Engineering (ASE)}, 2017.

\bibitem[Chen et~al.(2019)Chen, Kommrusch, Tufano, Pouchet, Poshyvanyk, and
  Monperrus]{sequencer}
Z.~Chen, S.~Kommrusch, M.~Tufano, L.~Pouchet, D.~Poshyvanyk, and M.~Monperrus.
\newblock Sequencer: Sequence-to-sequence learning for end-to-end program
  repair.
\newblock 2019.

\bibitem[Cheng et~al.(2020)Cheng, Ma, Zhang, and Xuan]{test-ibf20}
H.~Cheng, P.~Ma, J.~Zhang, and J.~Xuan.
\newblock Can this fault be detected by automated test generation: A
  preliminary study.
\newblock In \emph{2020 IEEE 2nd International Workshop on Intelligent Bug
  Fixing (IBF)}, 2020.

\bibitem[D'Antoni et~al.(2016)D'Antoni, Samanta, and Singh]{qclose}
L.~D'Antoni, R.~Samanta, and R.~Singh.
\newblock Qlose: Program repair with quantitative objectives.
\newblock In \emph{Computer Aided Verification}, 2016.

\bibitem[Durieux et~al.(2019)Durieux, Madeiral, Martinez, and
  Abreu]{RepairThemAll2019}
T.~Durieux, F.~Madeiral, M.~Martinez, and R.~Abreu.
\newblock {Empirical Review of Java Program Repair Tools: A Large-Scale
  Experiment on 2,141 Bugs and 23,551 Repair Attempts}.
\newblock In \emph{Proceedings of the 27th ACM Joint European Software
  Engineering Conference and Symposium on the Foundations of Software
  Engineering (ESEC/FSE '19)}, 2019.

\bibitem[Ernst et~al.(2007)Ernst, Perkins, Guo, McCamant, Pacheco, Tschantz,
  and Xiao]{daikon}
M.~Ernst, J.~Perkins, P.~Guo, S.~McCamant, C.~Pacheco, M.~Tschantz, and
  C.~Xiao.
\newblock The daikon system for dynamic detection of likely invariants.
\newblock \emph{Science of Computer Programming}, 69, 2007.

\bibitem[Ernst et~al.(1999)Ernst, Cockrell, Griswold, and Notkin]{invariant}
M.~D. Ernst, J.~Cockrell, W.~G. Griswold, and D.~Notkin.
\newblock Dynamically discovering likely program invariants to support program
  evolution.
\newblock In \emph{Proceedings of the 21st International Conference on Software
  Engineering}, ICSE ’99, New York, NY, USA, 1999. Association for Computing
  Machinery.

\bibitem[Experiment(2020)]{repo}
G.~Experiment.
\newblock Experiment repository is available on github:.
\newblock \emph{https://github.com/KTH/drr}, 2020.

\bibitem[Fisher et~al.(2011)Fisher, Wloka, Tip, Ryder, and
  Luchansky]{coverage-criteria}
M.~Fisher, J.~Wloka, F.~Tip, B.~G. Ryder, and A.~Luchansky.
\newblock An evaluation of change-based coverage criteria.
\newblock In \emph{Proceedings of the 10th ACM SIGPLAN-SIGSOFT Workshop on
  Program Analysis for Software Tools}. Association for Computing Machinery,
  2011.

\bibitem[Fraser and Arcuri(2011)]{evosuite}
G.~Fraser and A.~Arcuri.
\newblock Evosuite: automatic test suite generation for object-oriented
  software.
\newblock In \emph{ESEC/FSE '11 Proceedings of the 19th ACM SIGSOFT symposium
  and the 13th European conference on Foundations of software engineering},
  2011.

\bibitem[Fraser and Arcuri(2013)]{challenge-evosuite-icst13-Fraser}
G.~Fraser and A.~Arcuri.
\newblock Evosuite: On the challenges of test case generation in the real
  world.
\newblock In \emph{2013 IEEE Sixth International Conference on Software
  Testing, Verification and Validation}, 2013.

\bibitem[Gao et~al.(2019)Gao, Mechtaev, and Roychoudhury]{ISSTA19GaoCrash}
X.~Gao, S.~Mechtaev, and A.~Roychoudhury.
\newblock Crash-avoiding program repair.
\newblock In \emph{Proceedings of the 28th ACM SIGSOFT International Symposium
  on Software Testing and Analysis}, ISSTA 2019, 2019.

\bibitem[Gazzola et~al.(2017)Gazzola, Micucci, and Mariani]{TSE-repair-survey}
L.~Gazzola, D.~Micucci, and L.~Mariani.
\newblock Automatic software repair: A survey.
\newblock \emph{IEEE Transactions on Software Engineering}, 2017.

\bibitem[Ghanbari(2019)]{ghanbari2019validation}
A.~Ghanbari.
\newblock Validation of automatically generated patches: An appetizer.
\newblock Technical Report 1912.00117, arXiv, 2019.

\bibitem[Gu et~al.(2010)Gu, Barr, Hamilton, and Su]{Gu2010BRF}
Z.~Gu, E.~T. Barr, D.~J. Hamilton, and Z.~Su.
\newblock Has the bug really been fixed?
\newblock In \emph{Proceedings of the 32Nd ACM/IEEE International Conference on
  Software Engineering - Volume 1}, ICSE '10, New York, NY, USA, 2010. ACM.

\bibitem[{Hao} et~al.(2016){Hao}, {Zhang}, {Zang}, {Wang}, {Wu}, and
  {Xie}]{test-prio-3}
D.~{Hao}, L.~{Zhang}, L.~{Zang}, Y.~{Wang}, X.~{Wu}, and T.~{Xie}.
\newblock To be optimal or not in test-case prioritization.
\newblock \emph{IEEE Transactions on Software Engineering}, 42, May 2016.
\newblock ISSN 2326-3881.

\bibitem[Harrold et~al.(2000)Harrold, Rothermel, Sayre, Wu, and
  Yi]{SpectraRegression}
M.~Harrold, G.~Rothermel, K.~Sayre, R.~Wu, and L.~Yi.
\newblock An empirical investigation of the relationship between spectra
  differences and regression faults.
\newblock \emph{Software Testing, Verification and Reliability}, 2000.

\bibitem[{Hsu} and {Orso}(2009)]{test-minimisation2}
H.~{Hsu} and A.~{Orso}.
\newblock Mints: A general framework and tool for supporting test-suite
  minimization.
\newblock In \emph{2009 IEEE 31st International Conference on Software
  Engineering}, pages 419--429, May 2009.

\bibitem[Hua et~al.(2018)Hua, Zhang, Wang, and Khurshid]{sketchfix}
J.~Hua, M.~Zhang, K.~Wang, and S.~Khurshid.
\newblock Towards practical program repair with on-demand candidate generation.
\newblock In \emph{Proceedings of the 40th International Conference on Software
  Engineering}, 2018.

\bibitem[{Jeffrey} and {Neelam Gupta}(2005)]{test-minimisation1}
D.~{Jeffrey} and {Neelam Gupta}.
\newblock Test suite reduction with selective redundancy.
\newblock In \emph{21st IEEE International Conference on Software Maintenance
  (ICSM'05)}, pages 549--558, 2005.

\bibitem[Jiang et~al.(2018)Jiang, Xiong, Zhang, Gao, and Chen]{Simfix:2018}
J.~Jiang, Y.~Xiong, H.~Zhang, Q.~Gao, and X.~Chen.
\newblock Shaping program repair space with existing patches and similar code.
\newblock ISSTA, 2018.

\bibitem[Just et~al.(2014)Just, Jalali, and Ernst]{defect4j-dataset}
R.~Just, D.~Jalali, and M.~D. Ernst.
\newblock Defects4j: A database of existing faults to enable controlled testing
  studies for java programs.
\newblock In \emph{Proceedings of the 2014 International Symposium on Software
  Testing and Analysis}, ISSTA 2014, 2014.

\bibitem[Kifetew et~al.(2019)Kifetew, Devroey, and
  Rueda]{automatic-test-competition}
F.~Kifetew, X.~Devroey, and U.~Rueda.
\newblock Java unit testing tool competition: Seventh round.
\newblock In \emph{Proceedings of the 12th International Workshop on
  Search-Based Software Testing}, SBST ’19, 2019.

\bibitem[Lau and Yu(2005)]{hierarchy-minimisation}
M.~Lau and Y.~Yu.
\newblock An extended fault class hierarchy for specification-based testing.
\newblock \emph{ACM Transactions on Software Engineering and Methodology
  (TOSEM)}, 14:\penalty0 247--276, 07 2005.

\bibitem[Le et~al.(2016)Le, Lo, and Le~Goues]{hdrepair}
X.~B.~D. Le, D.~Lo, and C.~Le~Goues.
\newblock History driven program repair.
\newblock In \emph{Software Analysis, Evolution, and Reengineering (SANER),
  2016 IEEE 23rd International Conference on}, volume~1. IEEE, 2016.

\bibitem[Le et~al.(2017)Le, Chu, Lo, Le~Goues, and Visser]{s3}
X.-B.~D. Le, D.-H. Chu, D.~Lo, C.~Le~Goues, and W.~Visser.
\newblock S3: Syntax- and semantic-guided repair synthesis via programming by
  examples.
\newblock In \emph{Proceedings of the 2017 11th Joint Meeting on Foundations of
  Software Engineering}, ESEC/FSE 2017, 2017.

\bibitem[Le et~al.(2018)Le, Thung, Lo, and Goues]{Le:overfitting}
X.-B.~D. Le, F.~Thung, D.~Lo, and C.~L. Goues.
\newblock Overfitting in semantics-based automated program repair.
\newblock In \emph{Proceedings of the 40th International Conference on Software
  Engineering}, ICSE '18, New York, NY, USA, 2018. ACM.

\bibitem[Le et~al.(2019)Le, Bao, Lo, Xia, Li, and
  Pasareanu]{le:reliability-patch-assess}
X.-B.~D. Le, L.~Bao, D.~Lo, X.~Xia, S.~Li, and C.~Pasareanu.
\newblock On reliability of patch correctness assessment.
\newblock In \emph{Proceedings of the 41st ACM/IEEE International Conference on
  Software Engineering}, 2019.

\bibitem[Liu and Zhong(2018)]{sofix}
X.~Liu and H.~Zhong.
\newblock Mining stackoverflow for program repair.
\newblock In \emph{2018 IEEE 25th International Conference on Software
  Analysis, Evolution and Reengineering (SANER)}, 2018.

\bibitem[Long and Rinard(2016)]{Long-search-space}
F.~Long and M.~Rinard.
\newblock An analysis of the search spaces for generate and validate patch
  generation systems.
\newblock In \emph{Proceedings of the 38th International Conference on Software
  Engineering}, ICSE '16, New York, NY, USA, 2016. ACM.

\bibitem[Madeiral et~al.(2019)Madeiral, Urli, Maia, and Monperrus]{Bears2019}
F.~Madeiral, S.~Urli, M.~Maia, and M.~Monperrus.
\newblock {Bears: An Extensible Java Bug Benchmark for Automatic Program Repair
  Studies}.
\newblock In \emph{Proceedings of the 26th IEEE International Conference on
  Software Analysis, Evolution and Reengineering (SANER '19)}, Hangzhou, China,
  2019. IEEE.

\bibitem[Mann and Whitney(1947)]{Mann-whitney-utest}
H.~B. Mann and D.~R. Whitney.
\newblock On a test of whether one of two random variables is stochastically
  larger than the other.
\newblock \emph{The Annals of Mathematical Statistics}, 1947.

\bibitem[Marinescu and Cadar(2012)]{ICSE12-Symoblic-for-regression}
P.~D. Marinescu and C.~Cadar.
\newblock Make test-zesti: A symbolic execution solution for improving
  regression testing.
\newblock In \emph{Proceedings of the 34th International Conference on Software
  Engineering}, ICSE ’12, page 716–726. IEEE Press, 2012.
\newblock ISBN 9781467310673.

\bibitem[Marinescu and Cadar(2013)]{katch-fse13}
P.~D. Marinescu and C.~Cadar.
\newblock Katch: High-coverage testing of software patches.
\newblock In \emph{European Software Engineering Conference / ACM SIGSOFT
  Symposium on the Foundations of Software Engineering (ESEC/FSE 2013)}, 8
  2013.

\bibitem[Martinez and Monperrus(2018)]{cardumen}
M.~Martinez and M.~Monperrus.
\newblock Ultra-large repair search space with automatically mined templates:
  the cardumen mode of astor.
\newblock In \emph{{SSBSE 2018 - 10th International Symposium on Search-Based
  Software Engineering}}, 2018.

\bibitem[Martinez et~al.(2016)Martinez, Durieux, Sommerard, Xuan, and
  Monperrus]{martias2016defects4j}
M.~Martinez, T.~Durieux, R.~Sommerard, J.~Xuan, and M.~Monperrus.
\newblock {Automatic Repair of Real Bugs in Java: A Large-Scale Experiment on
  the Defects4J Dataset}.
\newblock \emph{Springer Empirical Software Engineering}, 2016.

\bibitem[Miller et~al.(1990)Miller, Fredriksen, and So]{fuzz}
B.~P. Miller, L.~Fredriksen, and B.~So.
\newblock An empirical study of the reliability of unix utilities.
\newblock \emph{Commun. ACM}, 1990.

\bibitem[Monperrus(2017)]{martinBibliography}
M.~Monperrus.
\newblock Automatic software repair: A bibliography.
\newblock \emph{ACM Comput. Surv.}, 51\penalty0 (1), Jan. 2017.
\newblock ISSN 0360-0300.

\bibitem[Orso et~al.(2004)Orso, Shi, and Harrold]{test-selection-2}
A.~Orso, N.~Shi, and M.~J. Harrold.
\newblock Scaling regression testing to large software systems.
\newblock In \emph{SIGSOFT '04/FSE-12}, 2004.

\bibitem[Pacheco and Ernst(2007)]{randoop}
C.~Pacheco and M.~D. Ernst.
\newblock Randoop: Feedback-directed random testing for java.
\newblock In \emph{Companion to the 22Nd ACM SIGPLAN Conference on
  Object-oriented Programming Systems and Applications Companion}, OOPSLA '07,
  2007.

\bibitem[Person et~al.(2008)Person, Dwyer, Elbaum, and
  Pundefinedsundefinedreanu]{diff-symbolic-exe}
S.~Person, M.~B. Dwyer, S.~Elbaum, and C.~S. Pundefinedsundefinedreanu.
\newblock Differential symbolic execution.
\newblock In \emph{Proceedings of the 16th ACM SIGSOFT International Symposium
  on Foundations of Software Engineering}, 2008.

\bibitem[Prasetya(2015)]{T3}
I.~S. W.~B. Prasetya.
\newblock T3i: A tool for generating and querying test suites for java.
\newblock In \emph{Proceedings of the 2015 10th Joint Meeting on Foundations of
  Software Engineering}, ESEC/FSE 2015. Association for Computing Machinery,
  2015.

\bibitem[Qi et~al.(2015)Qi, Long, Achour, and Rinard]{QiIssta15-overfitting}
Z.~Qi, F.~Long, S.~Achour, and M.~Rinard.
\newblock An analysis of patch plausibility and correctness for
  generate-and-validate patch generation systems.
\newblock In \emph{Proceedings of the 2015 International Symposium on Software
  Testing and Analysis}, ISSTA 2015, 2015.

\bibitem[Reps et~al.(1999)Reps, Ball, and Das]{PathSpectra}
T.~Reps, T.~Ball, and M.~Das.
\newblock The use of program profiling for software maintenance with
  applications to the year 2000 problem.
\newblock \emph{Software Engineering Notes}, 22, 1999.

\bibitem[Saha et~al.(2017)Saha, Lyu, Yoshida, and Prasad]{elixir}
R.~K. Saha, Y.~Lyu, H.~Yoshida, and M.~R. Prasad.
\newblock Elixir: Effective object oriented program repair.
\newblock In \emph{Proceedings of the 32Nd IEEE/ACM International Conference on
  Automated Software Engineering}, ASE 2017, 2017.

\bibitem[Shamshiri et~al.(2015)Shamshiri, Just, Rojas, Fraser, McMinn, and
  Arcuri]{ase15-tests}
S.~Shamshiri, R.~Just, J.~M. Rojas, G.~Fraser, P.~McMinn, and A.~Arcuri.
\newblock Do automatically generated unit tests find real faults? an empirical
  study of effectiveness and challenges.
\newblock In \emph{Proceedings of the 30th IEEE/ACM International Conference on
  Automated Software Engineering (ASE)}, 2015.

\bibitem[{Sherriff} et~al.(2007){Sherriff}, {Lake}, and
  {Williams}]{test-prio-2}
M.~{Sherriff}, M.~{Lake}, and L.~{Williams}.
\newblock Prioritization of regression tests using singular value decomposition
  with empirical change records.
\newblock In \emph{The 18th IEEE International Symposium on Software
  Reliability (ISSRE '07)}, pages 81--90, Nov 2007.

\bibitem[{Shriver} et~al.(2017){Shriver}, {Elbaum}, and
  {Stolee}]{Shriver-inputs}
D.~{Shriver}, S.~{Elbaum}, and K.~T. {Stolee}.
\newblock At the end of synthesis: Narrowing program candidates.
\newblock In \emph{2017 IEEE/ACM 39th International Conference on Software
  Engineering: New Ideas and Emerging Technologies Results Track (ICSE-NIER)},
  2017.

\bibitem[Smith et~al.(2015)Smith, Barr, Le~Goues, and Brun]{cure-worse-15}
E.~K. Smith, E.~T. Barr, C.~Le~Goues, and Y.~Brun.
\newblock Is the cure worse than the disease? overfitting in automated program
  repair.
\newblock In \emph{Proceedings of the 2015 10th Joint Meeting on Foundations of
  Software Engineering}, ESEC/FSE 2015, 2015.

\bibitem[Tan et~al.(2016)Tan, Yoshida, Prasad, and Roychoudhury]{anti-pattern}
S.~H. Tan, H.~Yoshida, M.~R. Prasad, and A.~Roychoudhury.
\newblock Anti-patterns in search-based program repair.
\newblock In \emph{Proceedings of the 2016 24th ACM SIGSOFT International
  Symposium on Foundations of Software Engineering}, FSE 2016, 2016.

\bibitem[Tonella(2004)]{etoc}
P.~Tonella.
\newblock Evolutionary testing of classes.
\newblock \emph{SIGSOFT Softw. Eng. Notes}, 2004.

\bibitem[Wang et~al.(2019)Wang, Wen, Chen, Yi, and Mao]{Shangwen19ESEM}
S.~Wang, M.~Wen, L.~Chen, X.~Yi, and X.~Mao.
\newblock How different is it between machine-generated and developer-provided
  patches? : An empirical study on the correct patches generated by automated
  program repair techniques.
\newblock In \emph{International Symposium on Empirical Software Engineering
  and Measurement}, 2019.

\bibitem[Wen et~al.(2018)Wen, Chen, Wu, Hao, and Cheung]{capgen-ICSE18}
M.~Wen, J.~Chen, R.~Wu, D.~Hao, and S.-C. Cheung.
\newblock Context-aware patch generation for better automated program repair.
\newblock In \emph{Proceedings of the 40th International Conference on Software
  Engineering}, ICSE '18, 2018.

\bibitem[White et~al.(2018)White, Tufano, Martinez, Monperrus, and
  Poshyvanyk]{deeprepair}
M.~White, M.~Tufano, M.~Martinez, M.~Monperrus, and D.~Poshyvanyk.
\newblock Sorting and transforming program repair ingredients via deep learning
  code similarities.
\newblock 2018.

\bibitem[Xie(2006)]{xietao}
T.~Xie.
\newblock Augmenting automatically generated unit-test suites with regression
  oracle checking.
\newblock In D.~Thomas, editor, \emph{ECOOP 2006 -- Object-Oriented
  Programming}, 2006.

\bibitem[Xie and Notkin(2004)]{prog-behav-diff-Xie}
T.~Xie and D.~Notkin.
\newblock Checking inside the black box: regression testing based on value
  spectra differences.
\newblock In \emph{20th IEEE International Conference on Software Maintenance,
  2004. Proceedings.}, pages 28--37, Sep. 2004.

\bibitem[Xin and Reiss(2017{\natexlab{a}})]{issta17-difftgen}
Q.~Xin and S.~Reiss.
\newblock Identifying test-suite-overfitted patches through test case
  generation.
\newblock In \emph{Proceedings of the 26th ACM SIGSOFT International Symposium
  on Software Testing and Analysis}, ISSTA 2017, page 226–236, New York, NY,
  USA, 2017{\natexlab{a}}. Association for Computing Machinery.
\newblock ISBN 9781450350761.

\bibitem[Xin and Reiss(2017{\natexlab{b}})]{ssFix}
Q.~Xin and S.~Reiss.
\newblock Leveraging syntax-related code for automated program repair.
\newblock In \emph{2017 32nd IEEE/ACM International Conference on Automated
  Software Engineering (ASE)}, 2017{\natexlab{b}}.

\bibitem[Xiong et~al.(2017)Xiong, Wang, Yan, Zhang, Han, Huang, and
  Zhang]{Xiong-ACS-ICSE17}
Y.~Xiong, J.~Wang, R.~Yan, J.~Zhang, S.~Han, G.~Huang, and L.~Zhang.
\newblock Precise condition synthesis for program repair.
\newblock In \emph{Proceedings of the 39th International Conference on Software
  Engineering}, 2017.

\bibitem[Xiong et~al.(2018)Xiong, Liu, Zeng, Zhang, and
  Huang]{XiongIdentifyingPC-ICSE18-patchsim}
Y.~Xiong, X.~Liu, M.~Zeng, L.~Zhang, and G.~Huang.
\newblock Identifying patch correctness in test-based program repair.
\newblock In \emph{Proceedings of the 40th International Conference on Software
  Engineering}, 2018.

\bibitem[{Xu} and {Rountev}(2007)]{test-selection-3}
G.~{Xu} and A.~{Rountev}.
\newblock Regression test selection for aspectj software.
\newblock In \emph{29th International Conference on Software Engineering
  (ICSE'07)}, 2007.

\bibitem[Yang et~al.(2017)Yang, Zhikhartsev, Liu, and Tan]{Opad}
J.~Yang, A.~Zhikhartsev, Y.~Liu, and L.~Tan.
\newblock Better test cases for better automated program repair.
\newblock In \emph{In Proceedings of 2017 11th Joint Meeting of the European
  Software Engineering Conference and the ACM SIGSOFT Symposium on the
  Foundations of Software Engineering, Paderborn, Germany, September 4–8,
  2017 (ESEC/FSE’17)}, 2017.

\bibitem[Ye et~al.(2020)Ye, Martinez, Durieux, and Monperrus]{quixbugs}
H.~Ye, M.~Martinez, T.~Durieux, and M.~Monperrus.
\newblock A comprehensive study of automatic program repair on the quixbugs
  benchmark.
\newblock \emph{Journal of Systems and Software}, 171:\penalty0 110825, 2020.
\newblock ISSN 0164-1212.
\newblock \doi{https://doi.org/10.1016/j.jss.2020.110825}.
\newblock URL
  \url{http://www.sciencedirect.com/science/article/pii/S0164121220302193}.

\bibitem[Yin et~al.(2011)Yin, Yuan, Zhou, Pasupathy, and
  Bairavasundaram]{Yin-fse11}
Z.~Yin, D.~Yuan, Y.~Zhou, S.~Pasupathy, and L.~Bairavasundaram.
\newblock How do fixes become bugs?
\newblock In \emph{Proceedings of the 19th ACM SIGSOFT Symposium and the 13th
  European Conference on Foundations of Software Engineering}, ESEC/FSE '11,
  2011.

\bibitem[Yoo and Harman(2007)]{test-selection-4}
S.~Yoo and M.~Harman.
\newblock Pareto efficient multi-objective test case selection.
\newblock In \emph{Proceedings of the 2007 International Symposium on Software
  Testing and Analysis}, ISSTA ’07. Association for Computing Machinery,
  2007.
\newblock ISBN 9781595937346.

\bibitem[Yoo and Harman(2012)]{regressionSurvey}
S.~Yoo and M.~Harman.
\newblock Regression testing minimization, selection and prioritization: A
  survey.
\newblock \emph{Software Testing, Verification and Reliability}, 22, 03 2012.
\newblock \doi{10.1002/stvr.430}.

\bibitem[Yu et~al.(2018)Yu, Martinez, Danglot, Durieux, and
  Monperrus]{zhongxing-EMSE18}
Z.~Yu, M.~Martinez, B.~Danglot, T.~Durieux, and M.~Monperrus.
\newblock Alleviating patch overfitting with automatic test generation: a study
  of feasibility and effectiveness for the nopol repair system.
\newblock \emph{Empirical Software Engineering}, 2018.

\bibitem[Yuan and Banzhaf(2018)]{Yuan2017ARJAAR}
Y.~Yuan and W.~Banzhaf.
\newblock Arja: Automated repair of java programs via multi-objective genetic
  programming.
\newblock In \emph{IEEE Transactions on Software Engineering}, 2018.

\bibitem[Zeller and Hildebrandt(2002)]{DeltaAlgorithm}
A.~Zeller and R.~Hildebrandt.
\newblock Simplifying and isolating failure-inducing input.
\newblock \emph{Software Engineering, IEEE Transactions on}, 2002.

\bibitem[Zhang(2018)]{hybrid-test-selection}
L.~Zhang.
\newblock Hybrid regression test selection.
\newblock In \emph{Proceedings of the 40th International Conference on Software
  Engineering}, ICSE ’18, page 199–209, New York, NY, USA, 2018.

\end{thebibliography}

\end{document}